\documentclass[journal]{IEEEtran}
\usepackage{arydshln}
\usepackage{graphicx}
\usepackage{booktabs}
\usepackage{multirow}
\usepackage{adjustbox}
\usepackage{amsmath}
\usepackage{amssymb}
\usepackage{pifont}
\newcommand{\cmark}{\ding{51}}%
\newcommand{\xmark}{\ding{55}}%
\usepackage{stackengine}

\usepackage{caption}
\usepackage{subcaption}
\usepackage{array}
\usepackage{cite}
\usepackage[dvipsnames]{xcolor}
\usepackage{hyperref}

\usepackage{times}
\usepackage{epsfig}
\usepackage{graphicx}
\usepackage{amsmath}
\usepackage{amssymb}
\usepackage{adjustbox}
\usepackage{booktabs}
\usepackage{graphicx}
\usepackage{booktabs}
\usepackage{multirow}
\usepackage{adjustbox}
\usepackage{amsmath}
\usepackage{amssymb}
\usepackage{pifont}
\usepackage{stackengine}
\usepackage{caption}
\usepackage{subcaption}
\usepackage{array}
\usepackage[dvipsnames]{xcolor}
\usepackage{tablefootnote}

\usepackage{arydshln}

\definecolor{diagram_green}{RGB}{130, 179, 102}
\definecolor{diagram_yellow}{RGB}{214, 182, 86}
\definecolor{diagram_red}{RGB}{184, 84, 80}
\definecolor{diagram_blue}{RGB}{108, 142, 191}
\definecolor{diagram_orange}{RGB}{215, 155, 0}
\definecolor{teaser_blue}{RGB}{37, 150, 190}

\ifCLASSINFOpdf
\else
\fi

\begin{document}
%
\title{Distilling Knowledge from Object Classification to Aesthetics Assessment}
%
%
%

\author{Jingwen Hou,
        Henghui Ding,
        Weisi Lin,~\IEEEmembership{Fellow,~IEEE,}
        Weide Liu,
        Yuming Fang,~\IEEEmembership{Senior Member,~IEEE}
\thanks{Jingwen Hou and Henghui Ding are co-first authors. Part of the work is done by Jingwen Hou during his internship at ByteDance.}
\thanks{Corresponding author: Weisi Lin.}}

%
%

\markboth{IEEE Transactions on XX,~Vol.~XX, No.~X, MONTH~YEAR}%
{Shell \MakeLowercase{\textit{et al.}}: Bare Demo of IEEEtran.cls for IEEE Journals}
%



\maketitle

\begin{abstract}

\textcolor{black}{
In this work, we point out that the major dilemma of image aesthetics assessment (IAA) comes from the abstract nature of aesthetic labels. That is, a vast variety of distinct contents can correspond to the same aesthetic label. On the one hand, during inference, the IAA model is required to relate various distinct contents to the same aesthetic label. On the other hand, when training, it would be hard for the IAA model to learn to distinguish different contents merely with the supervision from aesthetic labels, since aesthetic labels are not directly related to any specific content. To deal with this dilemma, we propose to distill knowledge on semantic patterns for a vast variety of image contents from multiple pre-trained object classification (POC) models to an IAA model. Expecting the combination of multiple POC models can provide sufficient knowledge on various image contents, the IAA model can easier learn to relate various distinct contents to a limited number of aesthetic labels. By supervising an end-to-end single-backbone IAA model with the distilled knowledge, the performance of the IAA model is significantly improved by 4.8\% in SRCC compared to the version trained only with ground-truth aesthetic labels. On specific categories of images, the SRCC improvement brought by the proposed method can achieve up to 7.2\%. Peer comparison also shows that our method outperforms 10 previous IAA methods.
}
\end{abstract}


%
\IEEEpeerreviewmaketitle

\section{Introduction}
\label{sec:intro}

\IEEEPARstart{I}{mage} aesthetics is significant in a variety of scenarios, including image recommendation \cite{10.1145/2647868.2655053}, image editing \cite{wang2018deep}, image retrieval \cite{obrador2009role}, and photo management \cite{li2010towards}. As a result, image aesthetics assessment (IAA) approaches are sought for evaluating visual aesthetic experiences automatically.
State-of-the-art (SOTA) methods \cite{lu2014rapid, lu2015deep, hosu2019effective, talebi2021learning, hou2020object} are mainly based on deep learning, which relies on neural networks for learning to extract aesthetic features (\textit{i.e.,} features for distinguishing different aesthetic levels) in a data-driven manner. 
As some works \cite{jin2019aesthetic, kao2017deep, zhang2020beyond} have suggested, semantic information can help to improve the effectiveness of a deep IAA model.  Kao \textit{et al.} \cite{kao2017deep} intuitively explained that semantic information is useful for IAA since humans need to understand the content of an image before assessing it. 
Besides such an intuitive explanation, we believe that the reason why semantic information is useful in IAA is that semantic information can make up for the shortcoming of the abstractness of aesthetic labels. 

\begin{figure}
\includegraphics[width=0.5\textwidth]{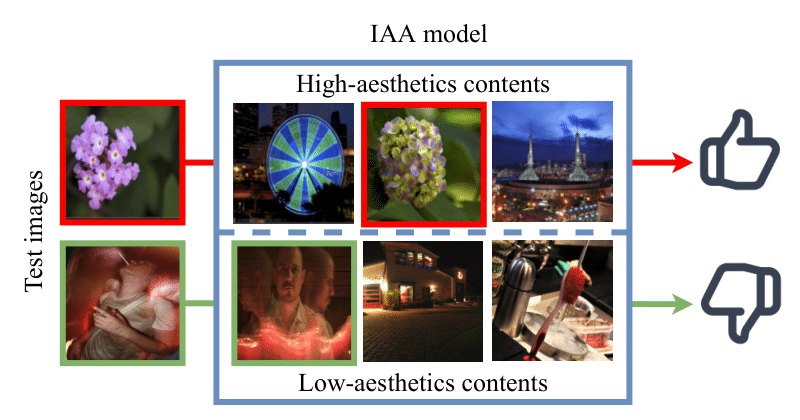}
\caption{Motivation for distilling knowledge from POC models. IAA by a deep model can be depicted as a process that maps contents of a test image into contents corresponding to different aesthetic levels known to the IAA model. In the given example, image pairs with boundaries of the same color represent relevant mappings. Since POC models can capture semantic patterns for a large variety of image contents, we distill knowledge from POC models to teach an IAA model on semantic patterns for relating a larger variety of contents to a limited number of aesthetic labels.
}
\label{fig:new_teaser}
\vspace{-6mm}
\end{figure}

\textcolor{black}{
Essentially, IAA can be regarded as a process that maps different image contents into different aesthetic levels (as shown in Fig. \ref{fig:new_teaser}). And image contents relevant to aesthetics are described by aesthetic features, which represent different combinations of semantic patterns relevant to IAA. 
For semantic patterns, we refer to a collection of pixels that are organized in a certain way so that such a pattern can be clearly identified across different images, and the combination of such patterns is connected to a certain semantic meaning (\textit{e.g.,} an object). 
And constructing discriminative aesthetic features requires sufficiently diverse semantic patterns to deal with a vast variety of contents. 
In the example of Fig. \ref{fig:new_teaser}, if the IAA model does not know the semantic patterns for representing a macro-photo of a flower, the IAA model cannot confidently map the test image of a flower to the class of high-aesthetics.
However, since similar aesthetic labels can refer to images with various contents, it is hard for deep models to learn semantic patterns from aesthetic labels.
To make up for the handicap of aesthetic labels in providing semantic guidance (\textit{i.e.,} guiding the IAA model to learn about semantic patterns), one instant way is to define predicting semantic information as an auxiliary task \cite{jin2019aesthetic, kao2017deep}, or use semantic information as an auxiliary input \cite{zhang2020beyond}. However, these approaches require extra human labels describing the semantics contained in images. 
}


\textcolor{black}{
Thus, we aim to provide semantic guidance to improve an IAA model without using labels from humans. 
Specifically, for an IAA model that produces less discriminative aesthetic features, we wish to provide extra supervision without human labels to guide the IAA model to capture more relevant semantic patterns for constructing more discriminative aesthetic features that can deal with a large variety of contents.
Typically, we consider a baseline IAA model constructed with a single pre-trained backbone trained merely with aesthetic labels.
Therefore, the aesthetic features of the IAA model are mostly constructed from the semantic patterns known to its selected pre-trained backbone, if its backbone fail to learn extra semantic patterns from aesthetic labels (and this is the case when training an IAA model merely with aesthetic labels, see Table \ref{tab:expq2} and Sec. \ref{sec:q2} for details). In this case, we may introduce extra pre-trained object classification (POC) models besides the selected pre-trained backbone to provide semantic guidance so that the IAA model can learn extra semantic patterns from them. 
Thus, one possible solution that provides semantic guidance to an IAA model with extra POC models is to encourage the IAA model to produce extra features as the extra POC models, which will force the IAA model to capture extra semantic patterns to construct extra features as the extra POC models. 
However, this cannot guarantee that the extra semantic patterns learned by the IAA model are relevant to the downstream IAA task, since POC models are trained on object classification instead of IAA.
}

\begin{figure}
\centering
\includegraphics[width=0.5\textwidth]{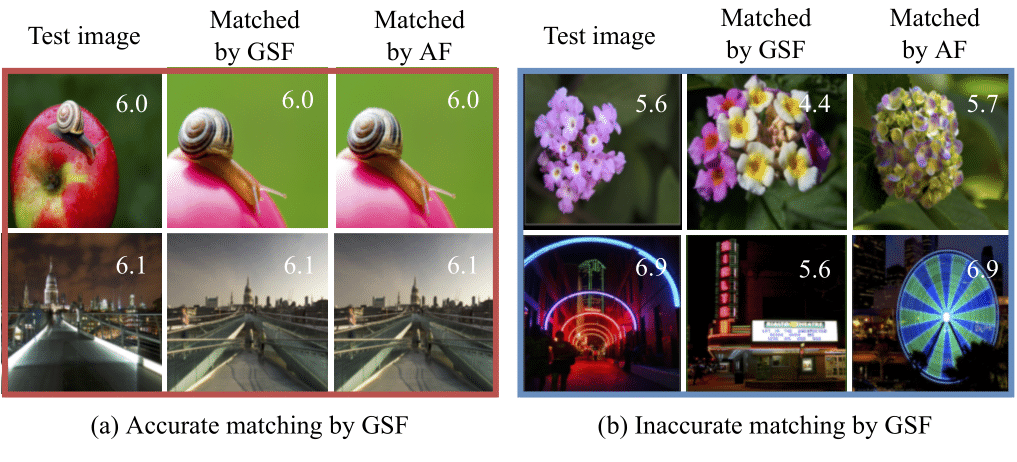}
\caption{
Comparing generic semantic features (GSFs) to aesthetic features (AFs) constructed from GSFs. The score for the test image is estimated by matching it to the most similar training image with GSFs or AFs. GT scores are given in the top-right corner. GSFs are likely to match visually-similar images with distinct scores, implying that some semantic patterns in GSFs may not be relevant to IAA. Thus, we take knowledge on IAA-relevant semantic patterns from GSFs instead of directly using GSFs for semantic guidance. Quantitative evaluations of GSFs and AFs on IAA by matching are given in Table \ref{tbl:GSF_vs_AF}.}
\label{fig:GSF_vs_AF}
\vspace{-6mm}
\end{figure}

\textcolor{black}{
For simplicity, we call features from the backbone of a POC model as generic semantic features (GSFs). We give an example in Fig. \ref{fig:GSF_vs_AF} to show whether GSFs can match images with similar aesthetic levels.
In some cases, images with similar GSFs can have similar aesthetic scores (as shown in Fig. \ref{fig:GSF_vs_AF}(a)). While in some other cases, though the GSFs can match images with similar visual appearances, the GSFs cannot fully distinguish images with different aesthetic levels (as shown in Fig. \ref{fig:GSF_vs_AF}(b)). This implies that some semantic patterns represented by GSFs may not be helpful for distinguishing aesthetic levels, but can even lead to confusion. 
Thus, we propose to use a knowledge distiller to select IAA-relevant semantic patterns from GSFs for constructing aesthetic features. Specifically, we train an IAA model as the knowledge distiller that can directly predict image aesthetics from the GSFs, and the inputs to the output layer of the knowledge distiller are regarded as aesthetic features. As shown in Fig. \ref{fig:GSF_vs_AF} (and Table \ref{tbl:GSF_vs_AF}), the aesthetic features constructed by the knowledge distiller can better distinguish different aesthetic levels than GSFs. 
Finally, the knowledge distiller can serve as the teacher model to provide a student IAA model with semantic guidance: the aesthetic features and predictions of the knowledge distiller are regarded as knowledge on semantic patterns distilled for IAA (\textit{i.e.,} teacher knowledge). The distilled knowledge is then used for imposing supervision to the student aesthetic features (\textit{i.e.,} input features to student's output layer) and student predictions. 
To ensure that the distilled knowledge can allow the student IAA model to acquire extra knowledge on semantic patterns for more discriminative aesthetic features, we choose POC models deeper or trained with more data than the student's backbone for constructing the teacher model, and verify that the teacher model has a higher IAA performance than the student.
}
Our contributions are summarized as follows:
\begin{itemize}
    \item \textcolor{black}{We point out the dilemma of IAA caused by the abstract nature of aesthetic labels. To deal with this dilemma, we propose a KD method to allow an end-to-end single-backbone IAA model can learn about semantic patterns relevant to IAA from multiple POC models via a knowledge distiller.}
    \item The knowledge distiller with the combined POC models as feature extractors is an effective IAA model (\textit{i.e.,} teacher model) which outperforms 10 previous IAA methods. Compared to the best-performed method among the contenders, the model achieves 5\% higher SRCC performance. 
    \item With the proposed KD scheme, the teacher model provides knowledge on semantic patterns to the training of a single-backbone end-to-end IAA model (\textit{i.e.,} student model). The performance of an end-to-end IAA model can be significantly improved by 4.8\% in SRCC compared to the version trained only with GT aesthetic labels. On specific categories of images, the improvement brought by the proposed KD scheme can achieve up to 7.2\%. Compared to the teacher model, the student model has a 99\% lower computational cost, with only 3\% lower SRCC performance. Peer comparison shows that the student model also outperforms 10 previous IAA methods. Compared to the best-performed end-to-end IAA model among the contenders, the student model achieves 7.1\% higher SRCC performance.
\end{itemize}

\section{Related Works}

\textcolor{black}{
\textbf{Semantic patterns in image aesthetics assessment (IAA)}. 
To build a robust IAA model, major efforts have been made to construct image features that distinguish different aesthetic levels (\textit{i.e.,} aesthetic features).
Assuming aesthetic levels can be distinguished by judging whether a photo follows known photography rules, early approaches attempt to predict image aesthetics from hand-crafted features following photography rules \cite{ sun2009photo, luo2011content, zhang2014fusion}. However, the number of well-defined photography rules is too limited to explain images on a large scale.
Therefore, SOTA methods \cite{lu2014rapid, lu2015deep, hosu2019effective, talebi2021learning, hou2020object} are mainly based on deep learning, which allows the model to learn to construct aesthetic features in a data-driven manner.
However, there is still not a consensus about what a deep IAA model has learned to distinguish images with different aesthetics. 
In this work, we hypothesize that IAA is a process that maps different combinations of semantic patterns (represented by aesthetic features) into different aesthetic levels. Thus, the IAA model is required to recognize more diverse relevant semantic patterns for aesthetic features when more diverse image contents are needed to be dealt with. 
This hypothesis can explain some findings in previous works: 1)  image aesthetics can be easier distinguished among images with similar GSFs \cite{tian2015query}, since it requires less semantic patterns to construct sufficiently discriminative aesthetic features; 2) image aesthetics can be accurately predicted  from GSFs \cite{hosu2019effective, hou2020object}, since GSFs contain semantic patterns useful for constructing aesthetic features; 3) introducing extra semantic information can help an IAA model to achieve higher performance \cite{jin2019aesthetic, kao2017deep, zhang2020beyond}, since it allows the IAA model to learn more semantic patterns for constructing aesthetic features. 
According to this hypothesis, we further propose methods to allow aesthetic features to be constructed with more diverse semantic patterns, so that the aesthetic features can deal with a larger variety of image contents: 1) we go beyond previous methods \cite{hosu2019effective, hou2020object} that construct aesthetic features with GSFs from a single POC model, we prepare more diverse semantic patterns with stacked GSFs from multiple POC models, which allows SOTA models to be achieved; 2) we also go beyond previous methods \cite{jin2019aesthetic, kao2017deep, zhang2020beyond} that introduce extra semantic labels, we distill knowledge on semantic patterns from multiple POC models to provide extra supervision to a weaker student model (\textit{e.g.,} a single-backbone end-to-end model), which allows us to significantly improve the performance of the student model without extra human efforts on labeling. 
}

\begin{figure*}[h]
\centering
\includegraphics[width=0.8\textwidth]{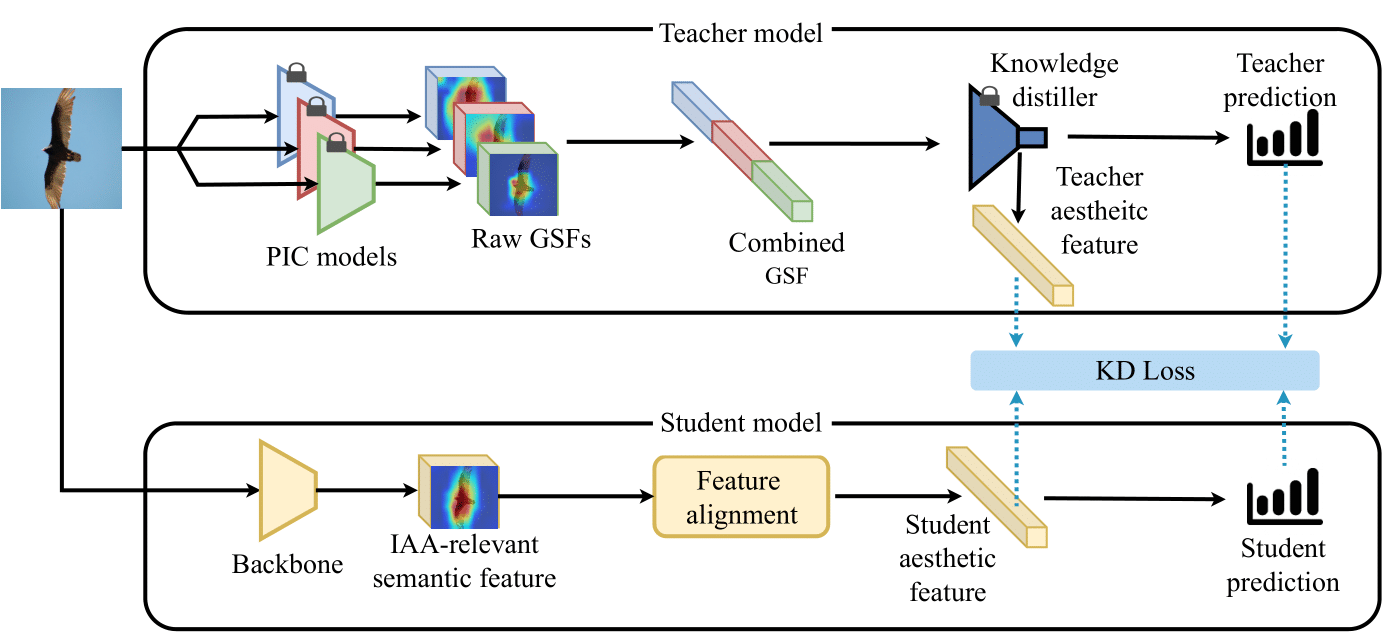}
\caption{Details of training a single-backbone end-to-end IAA model (\textit{i.e.,} student model) under the supervision of the teacher knowledge distilled from POC models via a knowledge distiller. The teacher aesthetic features and teacher predictions together are deemed as teacher knowledge distilled from GSFs. The teacher aesthetic features and teacher predictions are stored for imposing supervision on the aesthetic features and predictions of the student model when training the student model. Note that the preparation of teacher knowledge is conducted separately from the training of the student model in practice (see Table \ref{tab:cost} for the analysis on extra computational costs). }
\label{fig:overview}
\vspace{-6mm}
\end{figure*}

\textbf{Knowledge distillation (KD) beyond object classification.} KD has firstly been proposed by Hinton \textit{et al.} \cite{hinton2015distilling} for object classification, which transfers knowledge in the form of softened output logits from a deep teacher model to a rather shallower student model. Later on, various methods have been proposed for KD on object classification \cite{zhang2018deep, cho2019efficacy, romero2014fitnets, tung2019similarity, zhu2021complementary}. However, KD methods designed for object classification may not be fully applicable to other tasks. Therefore, various methods have been specifically designed for different tasks. 
In object detection, Chen \textit{et al.} \cite{chen2017learning} proposed to distill the knowledge on imbalanced classification and bounding box regression.
In semantic segmentation, He \textit{et al.} \cite{he2019knowledge} proposed to distill the knowledge on capturing long-term dependencies.
In road marking detection, Hou \textit{et al.} \cite{hou2020inter} proposed to distill the knowledge on the structural relationship of road scenes.  
\textcolor{black}{
In our case, we expect to improve an IAA model from knowledge on general object classification, where IAA and general object classification are two different tasks. As discussed, aesthetic features are essential for IAA performance, while more discriminative aesthetic features need to be built upon sufficiently diverse semantic patterns. To this end, we conduct KD to distill IAA-relevant knowledge on semantic patterns from object classification models. 
As a relevant topic, Zhang \textit{et al.}  \cite{zhang2021continual} has proposed to adopt a POC model for constructing an image quality assessment (IQA) model with continual learning that can deal with various IQA scenarios, which supports our hypothesis that GSFs can be used to guide the downstream IAA model to produce more discriminative features for distinguishing image aesthetics. 
}

\section{Our Approach}
\label{sec:approach}

\subsection{Problem Statement}
\label{sec:kd_iaa}
We consider a typical IAA model \cite{talebi2018nima} that estimates the aesthetic rating distribution directly from an image. Particularly, given the $i$-th image $\mathbf{I_i}$ from an IAA dataset, the IAA model $\mathcal{M}_{\theta}(\cdot)$ predicts the aesthetic rating distribution ${\mathcal{\hat D}_i}$:
\begin{equation}
    \mathcal{\hat D}_i = \mathcal{M}_\theta(\mathbf{I_i}).
\end{equation}
A direct way for obtaining the model $\mathcal{M}_{\theta}(\cdot)$ parameterized by $\theta$ is to directly optimize its parameter $\theta$ towards the ground-truth (GT) aesthetic rating distribution ${\mathcal{D}_i}$:
\begin{equation}
    \theta  = \mathop {\arg \min }\limits_\theta  \sum\limits_{i = 1}^N {\mathcal{L}({\mathcal{D}_i},{{\hat {\mathcal{D}}}_i})}, 
\end{equation}
where the most commonly-used loss function for $\mathcal{L}(\cdot)$ is earth mover distance (EMD) loss \cite{talebi2018nima}:
\begin{equation}
\label{eq:emd}
    {EMD({y},{{\hat {y}}})} = \sqrt { \frac{1}{n} \sum_{k=1}^{n}  \left | {\rm CDF}{_{{y}}}(k) - {\rm CDF}_{\hat {y}}(k)  \right |^2},
\end{equation}
where ${{\rm CDF}{_{{y}}}}$ and ${{\rm CDF}{_{{\hat y}}}}$ are cumulative density function for GT distribution $y$ and predicted distribution $\hat y$ of length $n$, respectively.

\textcolor{black}{
However, this approach overlooks the abstract nature of the aesthetic labels $\mathcal D_i$. 
As we previously discussed, IAA can be regarded as a process that maps different combinations of semantic patterns into different aesthetic levels. 
On the one hand, during inference, the IAA model is required to relate various combinations of semantic patterns to the same aesthetic label. On the other hand, when training, it would be hard for the IAA model to learn to distinguish different combinations of semantic patterns merely with the supervision from aesthetic labels since aesthetic labels are not directly related to any specific contents. 
To make up for the abstractness of aesthetic labels, one instant way is to assign an extra semantic label to each of the training samples for IAA. Because each semantic label is a direct description of the image contents (\textit{e.g.,} themes), the IAA model can learn about semantic patterns related to each semantic label along with the IAA objective, and aesthetic features covering more image contents can be constructed from the semantic patterns learned from semantic labels. }

\textcolor{black}{
Nevertheless, there are several problems with semantic labels: 1) it requires extra human efforts to assign semantic labels to each of IAA training samples; 2) it is hard to define what semantics in the image will be relevant to the downstream IAA task, and therefore, it is hard to find a standard for introducing semantic labels. Thus, our goal is to find a better representation that provides semantic guidance, so that the IAA model can learn about semantic patterns relevant to IAA for constructing more discriminative aesthetic features for a large variety of image contents: 
\begin{equation}
\label{eq:aesthetic_semantic}
    \theta  = \mathop {\arg \min }\limits_\theta  \sum\limits_{i = 1}^N ({\mathcal{L}_a({\mathcal{D}_i},{{\hat {\mathcal{D}}}_i}) + \mathcal{L}_s({\mathcal{S}_i},{{\hat {\mathcal{S}}}_i})}),
\end{equation}
where $\mathcal S_i$ and $\mathcal {\hat S}_i$ are the GT and predicted representation for semantic guidance, $\mathcal D_i$ and $\mathcal {\hat D}_i$ are the GT and predicted aesthetic rating distributions, and $\mathcal{L}_s(\cdot)$ and $\mathcal{L}_a(\cdot)$ are loss functions for semantic and aesthetic guidance, respectively. Here, semantic guidance is provided by imposing an extra supervision to the training objective.
}

\begin{figure}
\centering
\includegraphics[width=0.45\textwidth]{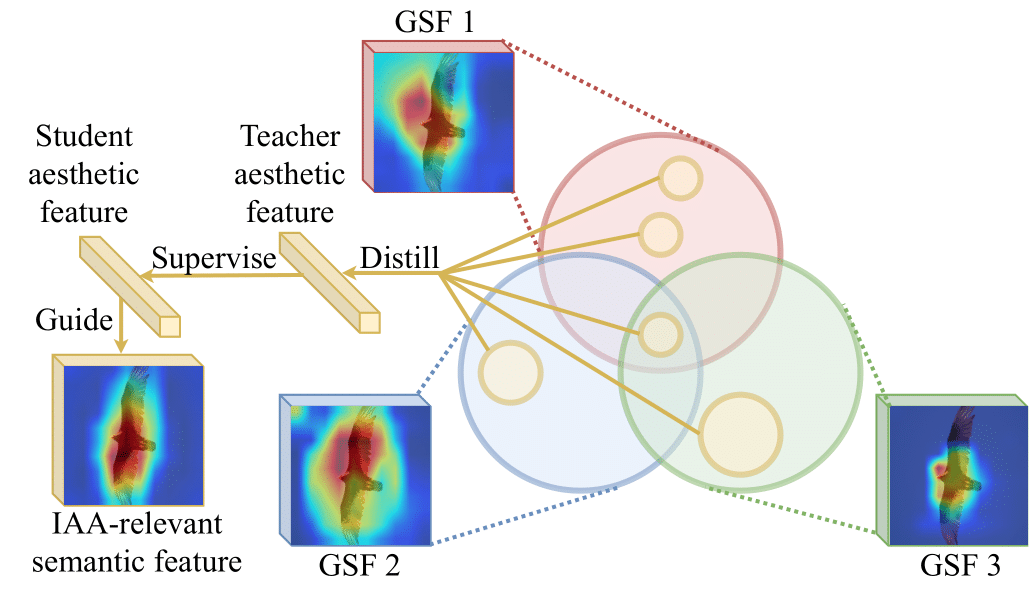}
\caption{LayerCAM \cite{jiang2021layercam} visualization for explaining how the proposed KD works in training the student model. For the same image, different GSFs tend to capture different semantic patterns. The knowledge distiller takes useful parts from GSFs to produce the teacher aesthetic feature. The teacher aesthetic features impose supervision to the student aesthetic features. Through back-propagation, the supervision from teacher aesthetic features further guides the student model to capture more IAA-relevant semantic features for producing more discriminative aesthetic features as the teacher. Details of training the knowledge distiller are given in Fig. \ref{fig:distiller}.}
\label{fig:redundant}
\vspace{-6mm}
\end{figure}

\subsection{Knowledge Distillation for Semantic Guidance}
\label{sec:feats}


As discussed in Sec.~\ref{sec:kd_iaa}, we aim for an extra supervision for semantic guidance besides extra semantic labels. 
Since POC models can recognize a vast variety of image contents, our idea is to distill knowledge from POC models on semantic patterns relevant to IAA. To cover a large variety of contents, we can take multiple POC models trained on different datasets to provide sufficiently diverse semantic patterns. And the semantic patterns can be taken from the GSFs of the selected POC models, as we earlier described in Sec. \ref{sec:intro}.
By combining GSFs from different POC models, we can obtain representations that can describe a large variety of contents:
\begin{equation}
\label{eq:gsf}
    f_C = [\bar f_1 | \bar f_2 | ... | \bar f_K],
\end{equation}
where $|$ denotes the concatentation operation that combines pooled GSFs $\bar f_k$ from $K$ different POC models, resulting in the combined feature $f_C$. For extracting features from different POC models, we adopt multi-layer spatial pooling (MLSP) \cite{hosu2019effective} as the pooling strategy to cover both low-level and high-level semantic information in the resulting feature, which is denoted as $\bar f_k = MLSP(f_k)$, where $f_k$ is the raw GSF from the $k$-th POC model.

However, directly supervising the IAA model by the combined GSFs is ineffective, since not all semantic patterns represented by the combined GSFs are relevant to IAA. We further describe the relationship between the combined GSFs and aesthetic features by \figurename~\ref{fig:redundant}. 
As shown, \textcolor{diagram_red}{red}, \textcolor{diagram_green}{green}, and \textcolor{diagram_blue}{blue} circles are sets of patterns that can be captured by different POC models. Since these models are trained with different datasets, different models may be sensitive to a different set of patterns. However, considering different POC models trained on different datasets may share similar categories of semantic patterns, these sets also have overlapped portions. 
For example, for classifying the same bird, some models may tend to classify by the beak, while some other models may tend to classify by wings. Nevertheless, it is possible that all models tend to classify an object as bird when they see the object with feather is in the sky. 
However, since these POC models are not trained for IAA, the patterns contained in these sets may not all be relevant to IAA. 
\textcolor{black}{
As we show in Fig. \ref{fig:GSF_vs_AF}, when GSFs contain semantic patterns not relevant to IAA, GSFs may match two visually-similar images but with distinct aesthetic scores.
In \figurename~\ref{fig:redundant}, we present the patterns relevant to IAA as \textcolor{diagram_yellow}{yellow}. }

\begin{figure}[t]
\centering
\includegraphics[width=0.45\textwidth]{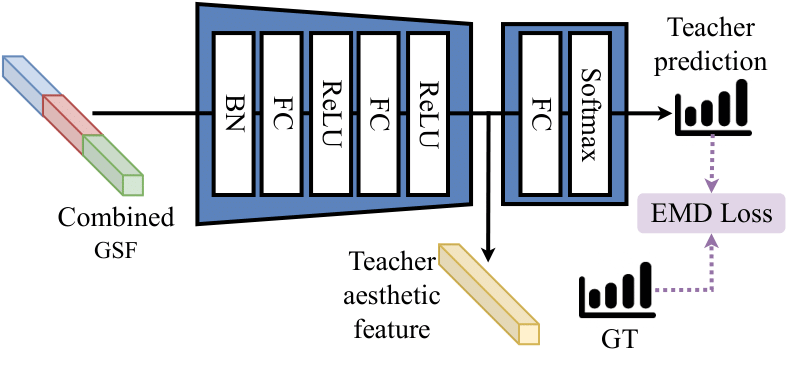}
\caption{Training of the knowledge distiller. Table \ref{tab:stacked_features} presents performance of the teacher IAA model constructed with the knowledge distiller and different POC models.}
\label{fig:distiller}
\vspace{-2mm}
\end{figure}

\begin{table}[t]
\caption{Comparison of GSFs for IAA from different models trained with ImageNet on the train-test split from  \cite{murray2012ava}. Full-resolution images are used.}
\centering
\renewcommand\arraystretch{1.1} 
\begin{tabular}{ccc|ccc}
\toprule
\multicolumn{3}{c|}{Model backbone} & \multicolumn{3}{c}{Results} \\
\midrule
ResNet18  & ResNet50  & ResNet101  & SRCC       & PLCC       & Acc       \\
\midrule
\cmark         &           &            & 0.669      & 0.678      & 78.7\%     \\
          & \cmark         &            & 0.713      & 0.718      & 80.2\%     \\
          &           & \cmark          & 0.720       & 0.724      & 80.3\%     \\
\cdashline{1-6}
\cmark         & \cmark         &            & 0.718      & 0.723      & 80.3\%     \\
          & \cmark         & \cmark          & 0.730       & 0.734      & 80.6\%     \\
\cmark         &           & \cmark          & 0.724      & 0.727      & 80.5\%     \\
\cdashline{1-6}
\cmark         & \cmark         & \cmark          & 0.730       & 0.734      & 80.7\%   \\ 
\bottomrule
\end{tabular}
\label{tbl:GSF_resnet}
\vspace{-4mm}
\end{table}

\begin{table}[t]
\caption{Comparison of GSFs for IAA from ResNeXt101 trained with different data on the train-test split from  \cite{murray2012ava}. Full-resolution images are used.}
\centering
\renewcommand\arraystretch{1.1} 
\begin{tabular}{ccc|ccc}
\toprule
\multicolumn{3}{c|}{Training setting} & \multicolumn{3}{c}{Results} \\
\midrule
ImageNet\tablefootnote{ResNeXt101\_32x8d: https://paperswithcode.com/lib/timm/resnext} & IG\tablefootnote{ig\_ResNeXt101\_32x8d: https://paperswithcode.com/lib/timm/ig-resnext} & SWSL\tablefootnote{swsl\_ResNeXt101\_32x8d: https://paperswithcode.com/model/swsl-resnext} & SRCC       & PLCC       & Acc       \\
\midrule
\cmark                     &                 &                   & 0.723      & 0.726      & 80.4\%      \\
                      & \cmark               &                   & 0.756      & 0.760       & 81.8\%     \\
                      &                 & \cmark                 & 0.755      & 0.758      & 81.5\%     \\
\cdashline{1-6}
\cmark                     & \cmark               &                   & 0.764      & 0.766      & 81.9\%     \\
                      & \cmark               & \cmark                 & 0.770       & 0.773      & 82.1\%     \\
\cmark                     &                 & \cmark                 & 0.763      & 0.765      & 81.9\%     \\
\cdashline{1-6}
\cmark                     & \cmark               & \cmark                 & 0.773      & 0.775      & 82.0\% \\     
\bottomrule
\end{tabular}
\label{tbl:GSF_resnext}
\vspace{-4mm}
\end{table}

Therefore, we propose a KD method based on a knowledge distiller trained separately. To be specific, we train an IAA model as the knowledge distiller with the combined GSF $f_C$. As shown in \figurename~\ref{fig:distiller}, the knowledge distiller is constructed with a batch normalization layer followed by three linear layers with ReLU activations. To train the knowledge distiller, we use EMD loss (Eq.~\ref{eq:emd}).
The unified whole of selected POC models and the trained knowledge distiller can also be viewed as a teacher model (upper-stream of \figurename~\ref{fig:overview}), and the single-backbone end-to-end model (lower-stream of \figurename~\ref{fig:overview}) is then viewed as the student model that learns to imitate the teacher. Thus, the knowledge distiller $\mathcal{C}(\cdot)$ predicts a feature $f_t$ and an aesthetic rating distribution $\mathcal{\hat D}_t$ from GSFs given by different POC models:
\begin{equation}
    \{f_t, \mathcal{\hat D}_t\} = \mathcal{C}(f_C), 
\end{equation}
where $\{f_t, \mathcal{\hat D}_t\}$ are deemed as teacher knowledge, including a teacher aesthetic feature $f_t$ and a teacher prediction $\mathcal{\hat D}_t$.


\begin{table*}[]
\caption{Comparison of different POC models as feature extractors of the teacher model. Experiments are conducted on the same train-test split from \cite{murray2012ava}, and results show that the knowledge distiller can effectively distill information needed for IAA from GSFs.}
\renewcommand\arraystretch{1.1} 
\begin{adjustbox}{width=2\columnwidth,center}
\begin{tabular}{c|ccc|ccc|ccc}
\toprule
\multirow{2}{*}{Setting} & \multicolumn{3}{c|}{Model backbone}                    & \multicolumn{3}{c|}{Full resolution ($\sim$640$\times$640)} & \multicolumn{3}{c}{Resized (300$\times$300)} \\ \cline{2-10}
                         & ResNet-v2 (BiTm)\tablefootnote{resnetv2\_152x4\_bitm: https://paperswithcode.com/lib/timm/big-transfer} & ResNeXt101 (SWSL)\tablefootnote{swsl\_ResNeXt101\_32x8d: https://paperswithcode.com/model/swsl-resnext} & ResNeXt101 (IG)\tablefootnote{ig\_ResNeXt101\_32x48d: https://paperswithcode.com/lib/timm/ig-resnext} & SRCC$\uparrow$       & PLCC$\uparrow$       & Acc$\uparrow$       & SRCC$\uparrow$       & PLCC$\uparrow$       & Acc$\uparrow$       \\
\midrule
1 & \cmark &   &   & 0.779 & 0.781 & 82.7\% & 0.762 & 0.764 & 82.2\% \\
2 &   & \cmark &   & 0.756 & 0.758 & 81.6\% & 0.739 & 0.742 & 80.8\% \\
3 &   &   & \cmark & 0.762 & 0.764 & 81.9\% & 0.753 & 0.754 & 81.6\% \\
\cdashline{2-10}
4 & \cmark & \cmark &   & 0.788 & 0.789 & 83.0\% & 0.771 & 0.773 & 82.3\% \\
5 &   & \cmark & \cmark & 0.775 & 0.777 & 82.2\% & 0.762 & 0.764 & 81.8\% \\
6 & \cmark &   & \cmark & 0.792 & 0.792 & 83.0\% & 0.779 & 0.779 & 82.4\% \\
\cdashline{2-10}
7 & \cmark & \cmark & \cmark & 0.794 & 0.795 & 83.1\% & 0.780 & 0.781 & 82.7\% \\
\bottomrule
\end{tabular}
\end{adjustbox}

\label{tab:stacked_features}
\vspace{-6mm}
\end{table*}

Thus, the teacher knowledge is directly used  for supervising the student model. Accordingly, we formulate the KD loss for training the student model from Eq. \ref{eq:aesthetic_semantic}:
\begin{equation}
\label{eq:kd_loss}
\mathcal{L_{KD}}({\mathcal{\hat D}_t}, {\mathcal{\hat D}_s}, {f_t}, {f_s}) = EMD({\mathcal{\hat D}_t}, {\mathcal{\hat D}_s}) + MSE({f_t}, {f_s}),
\end{equation}
where ${\mathcal{\hat D}_s}, {f_s}$ denote the student prediction and the student aesthetic feature, $EMD(\cdot)$ and $MSE(\cdot)$ refer to EMD loss and mean squared error (MSE) loss respectively.  
Accordingly, $f_t$ here is expected to provide semantic guidance in the context of IAA, and the student model is also expected to summarize how the teacher predicts $\mathcal {\hat D}_t$ from $f_t$. 
The architecture of the student model follows a succinct design (lower-stream of \figurename~\ref{fig:overview}),  which is constructed with a single CNN backbone followed by a fully-connected (FC) network for aligning the student aesthetic feature to the same size as the teacher aesthetic feature. The aesthetic labels are finally predicted from the student aesthetic feature by an FC softmax layer.

\textcolor{black}{
Note that in Eq. \ref{eq:kd_loss}, the teacher aesthetic feature $f_t$ is constructed from GSFs, and the student aesthetic feature $f_s$ is constructed from semantic patterns relevant to IAA captured by the student's own backbone.
And we hypothesize that the teacher aesthetic feature $f_t$ are more discriminative than the student aesthetic feature $f_s$ for effective semantic guidance. Generally, to prepare more discriminative teacher aesthetic features, the POC models for constructing the teacher model should be selected according to the student model. For a student model with a known backbone, we could expect the constituent POC models of the teacher model should: 1) deeper than the student's backbone; or 2) trained with more data than the student's backbone.
By combining multiple POC models deeper or trained with more data, more diverse semantic patterns are expected to be produced than the student's pre-trained backbone and more discriminative aesthetic features are expected to be created. 
\textbf{The points above are guidelines for selecting POC models that are likely to produce more diverse semantic patterns than the student's backbone. Note that since the effectiveness of a POC model also relies on the training strategy and the quality of the training data, we are not able to appropriately select POC models merely according to the parameter size and training data size.}
Thus, the criterion to verify whether the teacher aesthetic features are more discriminative than the student aesthetic features is to directly compare the performance of the teacher model (\textit{i.e.,} knowledge distiller) to the student model without KD. When the student model without KD performs poorer, it means that its aesthetic features are less discriminative than those of the teacher. As long as the teacher model performs better than the student model, the student model can learn to construct better aesthetic features from the teacher, and the selection of POC models for the teacher model is appropriate.
The points above will be experimentally discussed in Sec. \ref{sec:exp1} and Sec. \ref{sec:q2}. 
}

\begin{table}[t]
\centering
\caption{Comparison of generic semantic features (GSFs) and aesthetic features (AFs) on IAA by matching (as shown in Fig. \ref{fig:GSF_vs_AF}) on the split from \cite{murray2012ava}. Results show that AFs can better matches images of the same score.}
\begin{adjustbox}{width=0.6\columnwidth,center}
\begin{tabular}{l|lll}
\toprule
Feature & SRCC  & PLCC  & Acc   \\
\midrule
GSF     & 0.414 & 0.417 & 71.7\% \\
AF      & 0.657 & 0.657 & 78.3\% \\
\bottomrule
\end{tabular}
\end{adjustbox}
\label{tbl:GSF_vs_AF}
\vspace{-4mm}
\end{table}

\subsection{Adaptation to Smaller Input Sizes}
\label{sec:adapt_to_resize}

As shown in some previous works \cite{hosu2019effective, wang2019aspect}, resizing input images can harm the effectiveness of the IAA model due to loss in high-resolution details. However, using high-resolution images can introduce large computational costs. To cope with the trade-off between input resolution and computational costs, the proposed KD scheme is also designed to allow the student model to adapt to smaller input sizes. In training the teacher model, we adopt full-resolution inputs for feature extraction, which allows semantic patterns on high-resolution details to be preserved in the teacher knowledge. While for training the student model, we adopt resized inputs and encourage the student model to excavate high-resolution details from the resized image with the teacher knowledge. Experimental evidences are given in Table \ref{tab:expq23}.

\section{Experiments}
In this section, we would like to answer the following questions by experiments:
\begin{itemize}
\setlength\itemsep{0em}
\item How to construct more discriminative aesthetic features from GSFs? (Q1)
\item Does the supervision from the teacher model better than ground-truth aesthetic labels? (Q2)
\item How much improvement has been made in terms of efficiency, and how much effectiveness is compromised, by comparing the student to the teacher model?  (Q3)
\item How much improvement has been made by the teacher and the student model compared to previous works? (Q4)
\end{itemize}

\vspace{-4mm}
\subsection{Experimental Settings}
\label{sec:exp_settings}

\subsubsection{Dataset} Following previous works \cite{mai2016composition, lu2015deep, hosu2019effective, zhang2019gated, li2020personality, zeng2019unified, talebi2018nima, chen2020adaptive, she2021hierarchical}, our experiments are performed on the AVA dataset \cite{murray2012ava}. The AVA dataset includes $\sim$250,000 images, which have been scored from 1$\sim$10 by 78$\sim$594 workers. 
We follow previous works \cite{lu2015deep, mai2016composition, murray2017deep, zeng2019unified, zhang2019gated, hosu2019effective, hou2020object} to use the same train-test split\footnote{Note that the original source for the AVA dataset is no more available. The official split is originated from:  \url{https://github.com/mtobeiyf/ava\_downloader/blob/master/AVA\_dataset/aesthetics\_image\_lists/generic\_test.jpgl}} \cite{murray2012ava} for our experiments. The split adopts $\sim$230,000 images for training and $\sim$20,000 images for testing.

\subsubsection{Implementation details} Our model is implemented with PyTorch. 
As to the implementation of feature extraction, we refer to previous works \cite{yu2019beauty, hou2020attention, hosu2019effective}. 
For the teacher model, we train the model with batch size 512 for 12 epochs with Adam optimizer with an initial learning rate $3e\times10^{-5}$ and divided by 10 every 3 epochs. For the student model, we train the model with batch size 16 for 12 epochs with Adam optimizer with an initial learning rate $3e\times10^{-5}$ and divided by 10 every 3 epochs.

Following previous works \cite{talebi2018nima, zeng2019unified, zhang2019gated, hosu2019effective, hou2020object, she2021hierarchical}, we adopt Spearman Correlation Coefficients (SRCC), Pearson Correlation Coefficients (PLCC), and Accuracy (Acc) for evaluation. For evaluating SRCC and PLCC, we convert the aesthetic rating distributions into aesthetic scores by weighted average. As previous works \cite{mai2016composition, lu2015deep, hosu2019effective, zhang2019gated, li2020personality, zeng2019unified}, we take five as the cut-off threshold for converting aesthetic scores into binary aesthetic categories for evaluating Acc.  We also use Floating Point Operations (FLOPs)\footnote{\url{https://github.com/Swall0w/torchstat}} to evaluate computational costs.

\begin{table}[t]
\caption{Ablation study on the use of the proposed KD scheme on the same train-test split from \cite{murray2012ava}. The results imply that the proposed KD scheme enables the student's backbone to capture more IAA-relevant semantic patterns.}
\renewcommand\arraystretch{1.1} 
\begin{adjustbox}{width=1\columnwidth,center}
\begin{tabular}{cc|ccc}
\toprule
\multicolumn{2}{c|}{Setting}                                 & \multirow{2}{*}{SRCC$\uparrow$} & \multirow{2}{*}{PLCC$\uparrow$} & \multirow{2}{*}{Acc$\uparrow$} \\ \cline{1-2}
Trainable backbone & \multicolumn{1}{c|}{KD scheme} &                       &                       &                      \\
\midrule
\xmark                  & \xmark                                      & 0.736                 & 0.739                  & 81.1\%                \\
\cmark                  & \xmark                                      & 0.735                 & 0.737                 & 80.8\%              \\
\cdashline{1-5}
\xmark                  & \cmark                                      & 0.747                 & 0.748                  & 81.3\%                \\
\cmark                  & \cmark                                      & 0.770                 & 0.770                 & 82.1\%               \\
\bottomrule
\end{tabular}
\end{adjustbox}
\label{tab:expq2}
\end{table}

\begin{table}[t]
\caption{Ablation study on the two terms of the proposed KD loss (Eq. \ref{eq:kd_loss}) on the same train-test split from \cite{murray2012ava}. Results show that both terms of the KD loss contribute significantly to the overall improvement.}
\renewcommand\arraystretch{1.1} 
\begin{adjustbox}{width=0.77\columnwidth,center}
\begin{tabular}{cc|ccc}
\toprule
\multicolumn{2}{c|}{Supervision}       & \multicolumn{1}{c}{\multirow{2}{*}{SRCC$\uparrow$}} & \multicolumn{1}{c}{\multirow{2}{*}{PLCC$\uparrow$}} & \multicolumn{1}{c}{\multirow{2}{*}{Acc$\uparrow$}} \\ \cline{1-2}
Feature & \multicolumn{1}{l|}{Output} & \multicolumn{1}{c}{}                      & \multicolumn{1}{c}{}                      & \multicolumn{1}{c}{}                     \\
\midrule
\xmark       & \xmark                           & 0.735                                     & 0.737                                     & 80.8\%                                    \\
\cmark       & \xmark                           & 0.766                                     & 0.766                                     & 82.0\%                                    \\
\xmark       & \cmark                           & 0.758                                     & 0.758                                     & 81.6\%                                    \\
\cmark       & \cmark                           & 0.770                                     & 0.770                                     & 82.1\%                                   \\
\bottomrule
\end{tabular}
\end{adjustbox}
\label{tab:expq22}
\end{table}

\begin{table}[t]
\caption{Ablation study with smaller-sized inputs on the same train-test split from \cite{murray2012ava}. The performance drop brought by smaller input sizes is given in {\color{red}{red subscripts}}. The results show that model with KD has lower performance drop, implying that the proposed KD scheme enables the student model to adapt to smaller-sized inputs. }
\renewcommand\arraystretch{1.1} 
\begin{adjustbox}{width=1\columnwidth,center}
\begin{tabular}{c|l|lll|c}
\toprule
Input size & KD & SRCC$\uparrow$  & PLCC$\uparrow$  & Acc$\uparrow$ & FLOPs (G)$\downarrow$     \\
\midrule
640$\times$640          & \xmark  & 0.756 & 0.758 & 81.6\% &   134.7 \\
300$\times$300          & \xmark  & 0.736$\color{red}_{-2.6\%}$  & 0.739$\color{red}_{-2.5\%}$ & 81.1\% &   30.6 \\
\cdashline{1-6}
640$\times$640          & \cmark  & 0.775 & 0.777 & 82.4\% &   134.7 \\
300$\times$300          & \cmark  & 0.770$\color{red}_{-0.6\%}$ & 0.770$\color{red}_{-0.9\%}$ & 82.1\% &   30.6 \\
\bottomrule
\end{tabular}
\end{adjustbox}
\label{tab:expq23}
\vspace{-6mm}
\end{table}

\begin{table*}[]
\caption{Comparison between teacher and student models on the same train-test split from \cite{murray2012ava}. The results show that the student model sacrifices marginal effectiveness for a greater improvement in efficiency compared to the teachers.}
\renewcommand\arraystretch{1} 
\begin{adjustbox}{width=1.65\columnwidth,center}
\begin{tabular}{c|c|c|c|ccc}
\toprule
Model                                                                                     & Input Resolution & Params (M) & FLOPs (G) $\downarrow$ & SRCC $\uparrow$  & PLCC $\uparrow$  & Acc $\uparrow$   \\
\midrule
\begin{tabular}[c]{@{}c@{}}Teacher \\ Composite$^*$\end{tabular} & $\sim$640$\times$640          & 1853.2     & 2940.5    & 0.794 & 0.795 & 83.1\% \\
\midrule
\begin{tabular}[c]{@{}c@{}}Teacher \\ Composite$^*$\end{tabular} & 300$\times$300          & 1853.2     & 653.8     & 0.780 & 0.781 & 82.7\% \\
\midrule
\begin{tabular}[c]{@{}c@{}}Student\\ ResNeXt101 (SWSL) \end{tabular}                       & 640$\times$640          & 88.8      & 134.7     & 0.775 & 0.777 & 82.4\% \\
\midrule
\begin{tabular}[c]{@{}c@{}}Student\\ ResNeXt101 (SWSL) \end{tabular}                       & 300$\times$300          & 88.8     & 30.6      & 0.770 & 0.770 & 82.1\% \\
\bottomrule
\end{tabular}
\end{adjustbox}
\centering
\vspace{1mm} \\
$^*$: combines features from pre-trained ResNeXt101 (SWSL), ResNeXt101 (IG) and ResNetv2 (BiTm) \\
\label{tab:expq3}
\vspace{-4mm}
\end{table*}

\begin{table}[]
\caption{Comparison of results on different sub-categories of the end-to-end IAA model based on ResNeXt101(SWSL) with (\cmark) or without (\xmark) KD. The results are obtained from the split from \cite{murray2012ava}, and we choose sub-categories with top 8 portions in the testing set for the evaluation. The percentage of improvements brought by the proposed KD scheme is labeled in {\color{red}{red subscripts}}. The results show that the effectiveness of the proposed KD scheme is more obvious on some sub-categories.}
\renewcommand\arraystretch{1.1} 
\begin{adjustbox}{width=1\columnwidth,center}
\begin{tabular}{c|c|l|lll}
\toprule
\multicolumn{1}{l|}{Categories} & \multicolumn{1}{l|}{Portion} & KD   & SRCC$\uparrow$  & PLCC$\uparrow$  & Acc$\uparrow$ \\
\midrule
\multirow{2}{*}{Still life}         & \multirow{2}{*}{17.6\%}              & \xmark   & 0.695 & 0.706 & 77.7\%   \\
                               &                                      & \cmark   & 0.745$\color{red}_{+7.2\%}$ & 0.751$\color{red}_{+6.3\%}$ & 79.0\%   \\
\cdashline{1-6}
\multirow{2}{*}{Architecture}  & \multirow{2}{*}{15.5\%}              & \xmark    & 0.735 & 0.735 & 81.6\%   \\
                               &                                      & \cmark    & 0.761$\color{red}_{+3.6\%}$ & 0.759$\color{red}_{+3.2\%}$ & 82.7\%   \\
\cdashline{1-6}
\multirow{2}{*}{Landscape}     & \multirow{2}{*}{15.3\%}              & \xmark    & 0.777 & 0.776 & 83.8\%   \\
                               &                                      & \cmark    & 0.805$\color{red}_{+3.6\%}$ & 0.801$\color{red}_{+3.3\%}$ & 85.6\%   \\
\cdashline{1-6}
\multirow{2}{*}{Animals}       & \multirow{2}{*}{13.7\%}              & \xmark    & 0.734 & 0.734 & 80.4\%   \\
                               &                                      & \cmark    & 0.763$\color{red}_{+4.0\%}$ & 0.767$\color{red}_{+4.6\%}$ & 81.7\%   \\
\cdashline{1-6}
\multirow{2}{*}{Portraiture}   & \multirow{2}{*}{13.0\%}              & \xmark    & 0.689 & 0.693 & 82.8\%   \\
                               &                                      & \cmark    & 0.737$\color{red}_{+6.9\%}$ & 0.736$\color{red}_{+6.2\%}$ & 83.6\%   \\
\cdashline{1-6}
\multirow{2}{*}{Floral}        & \multirow{2}{*}{12.6\%}              & \xmark    & 0.742 & 0.738 & 79.0\%   \\
                               &                                      & \cmark    & 0.779$\color{red}_{+5.0\%}$ & 0.774$\color{red}_{+4.9\%}$ & 80.4\%   \\
\cdashline{1-6}
\multirow{2}{*}{Cityscape}     & \multirow{2}{*}{12.5\%}              & \xmark    & 0.749 & 0.749 & 81.6\%   \\
                               &                                      & \cmark    & 0.775$\color{red}_{+3.6\%}$ & 0.774$\color{red}_{+3.4\%}$ & 81.9\%   \\
\cdashline{1-6}
\multirow{2}{*}{Food}          & \multirow{2}{*}{12.5\%}              & \xmark    & 0.718 & 0.726 & 80.0\%   \\
                               &                                      & \cmark    & 0.764$\color{red}_{+6.4\%}$ & 0.771$\color{red}_{+6.1\%}$ & 82.0\%   \\
\midrule
\multirow{2}{*}{Overall}       & \multirow{2}{*}{100.0\%}             & \xmark    & 0.735 & 0.737 & 80.8\%   \\
                               &                                      & \cmark    & 0.770$\color{red}_{+4.8\%}$ & 0.770$\color{red}_{+4.6\%}$ & 82.1\%   \\
\bottomrule
\end{tabular}
\end{adjustbox}
\label{tab:subsets}
\vspace{-6mm}
\end{table}

\subsection{Effectiveness of Generic Semantic Features (Q1)}
\label{sec:exp1}

\textcolor{black}{
\subsubsection{Investigating aesthetic features constructed from GSFs} To answer the question: ``how to construct more discriminative aesthetic features from GSFs'', we have picked two sets of GSFs to investigate how GSFs from different POC models and their combinations will impact the IAA performance:
\begin{itemize}
    \item POC models of different architecture but trained on the same data: ResNet18, ResNet50 and ResNet101 trained on ImageNet are picked (see Table \ref{tbl:GSF_resnet}). 
    \item POC models of the same architecture but trained on different data: ResNeXt101 trained on different data are picked. Note that both IG version and SWSL version have included the ImageNet training data (see Table \ref{tbl:GSF_resnext}, sources for pre-trained models are given in the footnotes). 
\end{itemize} 
The approaches including using POC models with larger sizes or using POC models with more training data enable the GSFs to provide more diverse semantic patterns. Combining GSFs from different POC models further allows diverse semantic patterns from different POC models to be considered for aesthetic features.
Thus, the experimental results imply that the performance of different settings in Table \ref{tbl:GSF_resnet} and Table \ref{tbl:GSF_resnext} mainly depend on whether the semantic patterns are sufficiently diverse for constructing aesthetic features that can deal with diverse contents in the AVA dataset. 
Considering both the teacher model and the student model training merely with aesthetic labels construct aesthetic features from semantic patterns known to their pre-trained backbones (see Sec.\ref{sec:q2} for details), the teacher model consists of extra POC backbones with larger size and trained with more data, which is expected to capture more diverse semantic patterns for constructing aesthetic features covering more diverse contents. 
}


\textcolor{black}{
\subsubsection{Selecting POC models for KD} 
The targeted backbone for the student model we mainly consider is ResNeXt101 (SWSL) \cite{DBLP:journals/corr/abs-1905-00546}, whose baseline SRCC is 0.735 (Table \ref{tab:expq2}).
To construct the teacher model for later experiments in KD, we have selected three different POC models trained with different data, including ResNet-v2 (BiTm) \cite{kolesnikov2020big}, ResNeXt101 (IG) \cite{mahajan2018exploring}, and the same ResNeXt101 (SWSL) as the student (sources for the selected POC models are given in the footnotes of Table \ref{tab:stacked_features}). 
Since ResNet-v2 (BiTm) and ResNeXt101 (IG) are deeper or trained with more data than ResNeXt101 (SWSL), the combined GSFs are expected to contain semantic patterns captured by ResNeXt101 (SWSL) and extra semantic patterns captured by ResNet-v2 (BiTm) and ResNeXt101 (IG). 
The experimental results on teacher models using different combinations among the three POC models are given in Table \ref{tab:stacked_features}. The results for the best setting that combines all three POC models above have achieved 0.794 in SRCC. 
Since the performance is much higher than the student's, the selected POC models are valid for constructing the teacher model for the designated student based on ResNeXt101 (SWSL) according to the guidelines and criterion in Sec. \ref{sec:feats}.
In later sections, ``teacher model'' will refer to the teacher model trained with GSFs combined from ResNet-v2 (BiTm) , ResNeXt101 (IG) , and ResNeXt101 (SWSL) with SRCC 0.794.
Note that in the peer comparison of Sec. \ref{sec:q4}, we will also use the same teacher model for the student model based on ResNet18 or ResNet50 backbone pretrained on ImageNet. The selected POC models for the teacher model are all trained on ImageNet along with extra data, and are much deeper than ResNet18 and ResNet50. The baseline SRCC performance of ResNet18 and ResNet50 are 0.721 and 0.735 (Fig. \ref{fig:gain_variations}), which are poorer than the teacher's performance. Thus, using the same teacher model also follows the guidelines and criterion in Sec. \ref{sec:feats} and is expected to provide effective semantic guidance.
}

\textcolor{black}{
\subsubsection{Comparing GSFs to aesthetic features by matching} Based on the teacher model, we also compare the combined GSFs with the resulting aesthetic features following the matching-based approach as described in Fig. \ref{fig:GSF_vs_AF}. In this way, we are able to see whether the aesthetic features are more discriminative in aesthetics than the source GSFs. As the results shown in Table \ref{tbl:GSF_vs_AF}, the resulting aesthetic features significantly outperforms the source GSFs, which confirms the effectiveness of the proposed knowledge distiller.  
}



\begin{table*}[]
\caption{Comparison with reported results of SOTAs on the same train-test split from \cite{murray2012ava}. Best results are presented in {\color{blue}{blue}}, and the second best results are presented in {\color{red}{red}}. Publication years are given in subscripts.}
\renewcommand\arraystretch{1.1} 
\begin{adjustbox}{width=1.6\columnwidth,center}
\begin{tabular}{l|l|l|lll}
\toprule
Method                 & Backbone           & Input   size         & SRCC$\uparrow$   & PLCC$\uparrow$   & Acc$\uparrow$     \\
\midrule
DMA-Net$_{15}$ \cite{lu2015deep}               & AlexNet            & 224$\times$224              & -     & -     & 75.4\% \\
MNA-CNN$_{16}$ \cite{mai2016composition}               & VGG16              & 224$\times$224              & -     & -     & 77.1\% \\
NIMA$_{17}^\dag$ \cite{talebi2018nima}           & ResNet50           & 224$\times$224              & 0.654 & 0.662 & 78.6\% \\
APM$_{17}$ \cite{murray2017deep}                    & ResNet101          & 500$\times$500              & 0.709 & -     & 80.3\% \\
Hosu   \textit{et al.}$_{19}^\ddag$ \cite{hosu2019effective}          & InceptionResNet-v2 & 640$\times$640              & 0.756 & 0.757 & 81.7\% \\
Zeng   \textit{et al.}$_{19}$ \cite{zeng2019unified}            & ResNet101          & 384$\times$384              & 0.719 & 0.720 & 80.8\% \\
GPF-CNN$_{19}$ \cite{zhang2019gated}                & ResNet18           & 224$\times$224              & 0.671 & 0.682 & 80.3\% \\
Hou   \textit{et al.}$_{20}^\ddag$\cite{hou2020object}           & InceptionResNet-v2 & 640$\times$640 \&   320$\times$320 & 0.751 & 0.753 & 81.7\% \\
\midrule
Ours   (Teacher model$^\ddag$) & Composite$^*$     & $\sim$640$\times$640              & \color{blue}{0.794} & \color{blue}{0.795} & \color{blue}{83.1\%} \\
\cdashline{1-6}
Ours   (Student model) & ResNeXt101 (SWSL)         & 300$\times$300              & \color{red}{0.770} & \color{red}{0.770} & \color{red}{82.1\%} \\
Ours   (Student model) & ResNet50           & 300$\times$300              & 0.745 & 0.745 & 81.4\% \\
Ours   (Student model) & ResNet18           & 300$\times$300              & 0.719 & 0.722 & 80.5\% \\
\bottomrule
\end{tabular}
\end{adjustbox}
\label{tab:expq4_official}
\vspace{1mm} \\
\centering
$^*$: combines features from pre-trained ResNeXt101 (SWSL), ResNeXt101 (IG) and ResNetv2 (BiTm) \\
$^\dag$: we re-implement the model with ResNet50 and evaluate the model on the same train-test split as other presented methods.\\
$^\ddag$: not an end-to-end model
\end{table*}

\begin{table*}[]
\caption{Comparison with reported results of SOTAs on the same train-test split from \cite{jin2019ilgnet}. Best results are presented in {\color{blue}{blue}}, and the second best results are presented in {\color{red}{red}}. Publication years are given in subscripts.}
\renewcommand\arraystretch{1.1} 
\begin{adjustbox}{width=1.4\columnwidth,center}
\begin{tabular}{l|l|l|lll}
\toprule
Method                 & Backbone    & Input size    & SRCC$\uparrow$   & PLCC$\uparrow$   & Acc$\uparrow$     \\
\midrule
PA\_IAA$_{20}$ \cite{li2020personality}                & DenseNet121 & 299$\times$299       & 0.666 & -     & 82.9\% \\
HLA-GCN$_{21}$  \cite{she2021hierarchical}                 & ResNet101   & 300$\times$300       & 0.665 & 0.687 & 84.6\% \\
\midrule
Ours   (Teacher model) & Composite$^*$   & $\sim$640$\times$640 & \color{blue}{0.732} & \color{blue}{0.751} & \color{blue}{85.3\%} \\
\cdashline{1-6}
Ours   (Student model) & ResNeXt101 (SWSL)    & 300$\times$300       & \color{red}{0.701} & \color{red}{0.722} & \color{red}{84.9\%} \\
Ours   (Student model) & ResNet50    & 300$\times$300       & 0.677 & 0.698 & 84.1\% \\
Ours   (Student model) & ResNet18    & 300$\times$300       & 0.652 & 0.677 & 83.5\% \\
\bottomrule
\end{tabular}
\end{adjustbox}
\label{tab:expq4_ilgnet}
\vspace{1mm} \\
\centering
$^*$: combines features from pre-trained ResNeXt101 (SWSL), ResNeXt101 (IG) and ResNetv2 (BiTm) \\
\vspace{-4mm}
\end{table*}

\subsection{Effectiveness of Knowledge Distillation (Q2)}
\label{sec:q2}


To answer the question ``does the supervision from the teacher model better than ground-truth aesthetic labels", we conduct ablation studies to investigate different settings of the proposed KD scheme. 
Specifically, we firstly investigate whether a single backbone is capable of learning teacher knowledge and adapting to smaller input sizes under the proposed KD scheme. To this end, we set up baseline models trained directly with GT aesthetic rating distributions in the situation when the ResNeXt101 (SWSL) backbone is trainable or non-trainable.
Then we introduce the proposed KD scheme as comparisons. The results are given in Table \ref{tab:expq2}. 

\textcolor{black}{
As the results suggest, without the proposed KD scheme, similar results are obtained for both trainable and untrainable backbone settings. This suggests that even we give the backbone the freedom of learning on new semantic patterns for more discriminative aesthetic features, the IAA model has not been improved. 
This supports our motivation that aesthetic labels are too abstract to guide the neural network to learn about semantic patterns. 
As a result, the IAA model can only learn to create aesthetic features with semantic patterns already-known to its pre-trained backbone, and merely learn new semantic patterns for improving the discriminative power of the resulting aesthetic features. 
Since our teacher model also constructs aesthetic features from semantic patterns already-known to the backbones of the selected POC models, if we introduce extra POC models deeper or trained with more data than the pre-trained backbone of the student model,  it is expected that the teacher model will construct more discriminative aesthetic features than the student model. By encouraging the student aesthetic features to be close to the teacher aesthetic features in training, the student model is guided to capture more relevant semantic patterns for more discriminative aesthetic features as its teacher (Fig. \ref{fig:redundant}).
}

\textcolor{black}{
By introducing the proposed KD scheme, both trainable and untrainable backbone settings present substantial improvements in performance. The results have the following implications: 1) when the backbone is set untrainable, the student model with KD learns to construct more discriminative aesthetic features with semantic patterns already-known to its pre-trained backbone; 2) when the backbone is trainable, the backbone of the student model with KD learns relevant semantic patterns for more discriminative aesthetic features. This confirms the effectiveness of the proposed KD scheme. 
}

The KD loss (Eq. \ref{eq:kd_loss}) can be further divided into feature supervision (term 2 of Eq. \ref{eq:kd_loss}) and output supervision (term 1 of Eq. \ref{eq:kd_loss}). We detail our experiments to show the contribution of each individual term in the proposed KD loss. We set the whole network trainable in this experiment, and the model is directly supervised by GTs when the output supervision is off. The results are presented in Table \ref{tab:expq22}. As the results show, both terms substantially contribute to the performance of the student model, and the student model achieves the best performance when both terms are adopted.

Note that the student model takes 300$\times$300 resized inputs while it is supervised by teacher knowledge distilled from full-resolution inputs. As discussed in Sec. \ref{sec:adapt_to_resize}, this means we expect the student model to be adaptive to the smaller-sized inputs (learn full-resolution knowledge on semantic patterns from resized inputs). To confirm, we also train a student model with almost full-resolution (640$\times$640) inputs as a comparison. Specifically, we first pad all input images to 800$\times$800 (where 800$\times$800 is the largest size in the dataset), and center-crop them into 640$\times$640 to maintain the aspect ratio. The results are presented in Table \ref{tab:expq23}.
The results show that the proposed KD method also enables the student to be adaptive to smaller-sized inputs, which allows the performance drop brought by resizing (640$\times$640 $\rightarrow$ 300$\times$300) to be decreased from 2.6\% to 0.6\% in SRCC (0.756$\rightarrow$0.736 \textit{vs.} 0.775$\rightarrow$0.770), which leads to FLOPs to be saved by 77.2\% (134.7 $\rightarrow$ 30.6).
This suggests that the trained student model is adaptive to lower image resolution and change of aspect ratio after learning with teacher knowledge.

To show the impact of the proposed KD scheme on the performance of the student model on different categories of images, we further present the results on specific categories of the baseline model without KD and the student model with KD. The results are presented in Table \ref{tab:subsets}. As the results show, compared to the overall improvement 4.8\% on SRCC, the improvement brought by the proposed KD scheme can reach 7.2\% for specific categories. This further confirms the effectiveness of the proposed KD scheme.
%


\subsection{Comparing Student with Teacher Model (Q3)}
\label{sec:q3}

To answer the question ``how much improvement has been made in terms of efficiency, and how much effectiveness is compromised, by comparing the student to the teacher model", we compare the efficiency and effectiveness between teacher and student models. The results are presented in Table \ref{tab:expq3}. 
Compared to the teacher model, the student model with 640$\times$640 input size saves FLOPs by 95\% (2940.5 $\rightarrow$ 134.7) with 2.4\% drop in SRCC (0.794 $\rightarrow$ 0.775). By shrinking the input size to 300$\times$300, the FLOPs is further saved by 77.2\% (134.5 $\rightarrow$ 30.6) with only 0.6\% loss in SRCC (0.775$\rightarrow$0.770). 

\subsection{Comparison with State-of-the-arts (Q4)} 
\label{sec:q4}


To answer the question ``how much improvement has been made by the teacher and the student model compared to previous methods", we compare the teacher model and the student model with previous relevant methods 
 \cite{lu2015deep, mai2016composition, murray2017deep, talebi2018nima, zeng2019unified, zhang2019gated, hosu2019effective, hou2020object, li2020personality, she2021hierarchical}. We take their reported results for the comparison, and the results are presented in Table \ref{tab:expq4_official}. Among these contenders, the reported results of DMA-Net \cite{lu2015deep}, MNA-CNN \cite{mai2016composition},  Zeng \textit{et al.}'s method \cite{zeng2019unified}, APM \cite{murray2017deep}, Hou \textit{et al.}'s method \cite{hou2020object},  GPF-CNN \cite{zhang2019gated} and Hosu \textit{et al.}'s method \cite{hosu2019effective} are evaluated on the train-test split from \cite{murray2012ava}, while PA\_IAA \cite{li2020personality}  and HLA-GCN \cite{she2021hierarchical} are evaluated on the train-test split from \cite{jin2019ilgnet}\footnote{\url{https://github.com/BestiVictory/ILGnet/tree/local/data/AVA1}}. We experimentally observed that the results on the split from \cite{jin2019ilgnet} typically have a higher accuracy while a lower SRCC compared to the results on the split from  \cite{murray2012ava} (see more detailized discussions in Sec. \ref{sec:perform}). To fairly compare our method with the contenders, we separately compare the performance results of our method on the same train-test split as the contenders. Note that the reported results of NIMA have evaluated neither on the split from \cite{murray2012ava} nor on the split from \cite{jin2019ilgnet}. Therefore, we re-implement NIMA and evaluate on the split from \cite{murray2012ava}. \textbf{For the comparison on the split from \cite{murray2012ava}, the results are given in Table \ref{tab:expq4_official}; for the comparison on the split from \cite{jin2019ilgnet}, the results are given in Table \ref{tab:expq4_ilgnet}.}

\begin{figure*}[htb]
    \centering 
\begin{subfigure}{0.95\textwidth}
  \includegraphics[width=\linewidth]{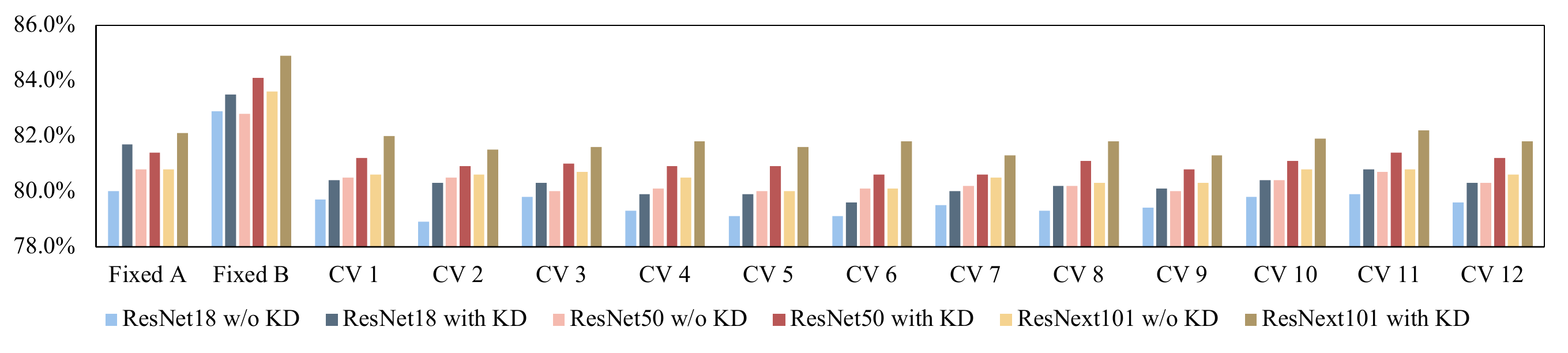}
  \vspace{-6mm}
  \caption{Accuracy results on fixed split evaluation and random split cross-validation with ResNet18 backbone. }
  \label{fig:1}
\end{subfigure}\hfil 
\begin{subfigure}{0.95\textwidth}
  \includegraphics[width=\linewidth]{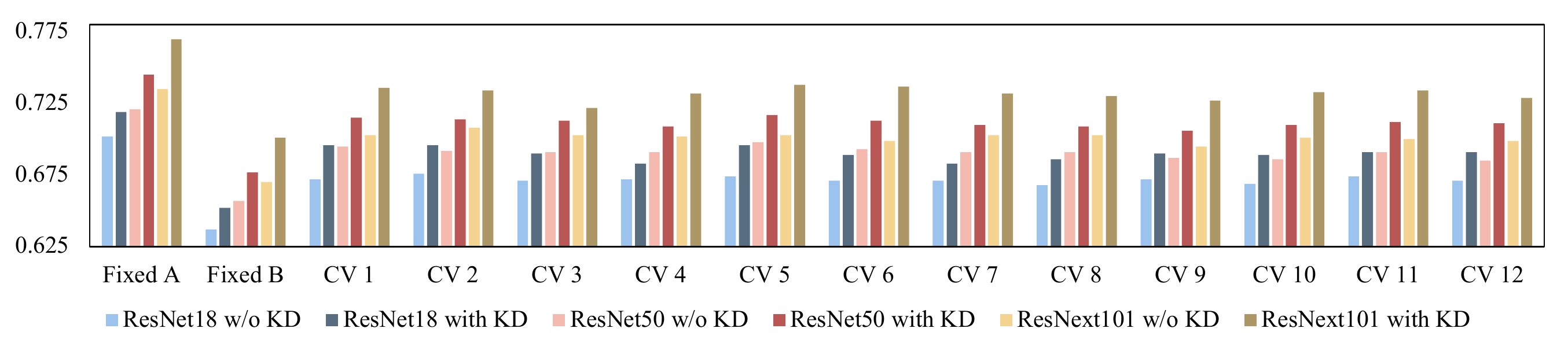}
  \vspace{-6mm}
  \caption{SRCC results on fixed split evaluation and random split cross-validation with ResNet18 backbone.}
  \label{fig:2}
\end{subfigure}\hfil 
\begin{subfigure}{0.95\textwidth}
  \includegraphics[width=\linewidth]{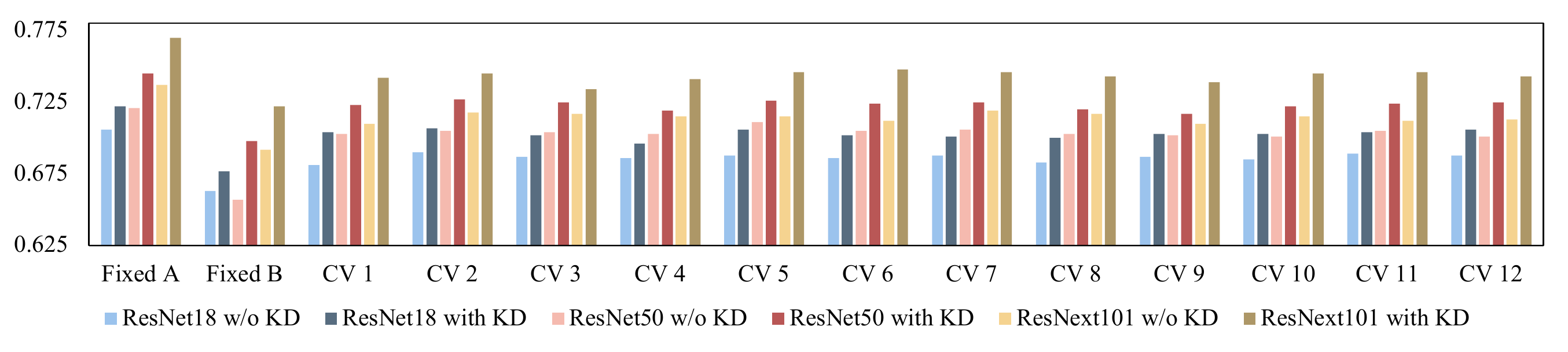}
  \vspace{-6mm}
  \caption{PLCC results on fixed split evaluation and random split cross-validation with ResNet18 backbone.}
  \label{fig:3}
\end{subfigure}\hfil 
\caption{Demonstration of generalizability of the proposed KD scheme across different train-test splits. Fixed A and Fixed B denote the train-test split from \cite{murray2012ava} and \cite{jin2019ilgnet}, respectively. CV1 $\sim$ CV12 denote twelve cross-validation splits with non-overlapping testing sets.}
\label{fig:gain_variations}
\vspace{-6mm}
\end{figure*}

For our implementation, the ResNeXt101-based setting adopts the SWSL version \cite{DBLP:journals/corr/abs-1905-00546}, and the ResNet50 and the ResNet18 based settings utilize ImageNet pre-trained versions\footnote{\url{https://github.com/Cadene/pretrained-models.pytorch}} as the student model backbones. 
As shown in the results, the teacher model based on combined GSFs significantly outperforms the selected contenders in terms of SRCC, PLCC and accuracy. Specifically, the relevant method by Hosu \textit{et al.} \cite{hosu2019effective} based on InceptionResNet-v2 MLSP features, which performs best in terms of SRCC and PLCC among all contenders, is substantially surpassed by the teacher model by 5\%.
Additionally, the proposed KD scheme helps to significantly improve the model efficiency while preserving the model effectiveness. 
As shown in Table \ref{tab:expq4_official} and Table \ref{tab:expq4_ilgnet}, though a performance drop can be observed comparing the teacher model to the student model, the student model based on ResNeXt101 still outperforms the contenders in terms of SRCC, PLCC and accuracy, including the most recent methods \cite{she2021hierarchical, li2020personality, hou2020object}.  Comparing the student model to other end-to-end models among the contenders, our student model achieves 0.770 in SRCC, which is 7.1\% higher than the best performed end-to-end models among the contenders (Zeng \textit{et al.}'s method \cite{zeng2019unified} in Table \ref{tab:expq4_official} ). 

\section{Further Discussion}

\begin{table*}[]
\caption{Comparison between the proposed KD scheme and other approaches for semantic guidance. Methods are evaluated on the split from \cite{murray2012ava}. Results show that the proposed KD scheme is more effective than using extra semantic labels.}
\renewcommand\arraystretch{1.1} 
\begin{adjustbox}{width=2\columnwidth,center}
\begin{tabular}{l|l|l|l|llll}
\toprule
Model                & Input                                                                             & Output                                                                                & Loss                                                                     & SRCC$\uparrow$  & PLCC$\uparrow$  & Acc$\uparrow$    \\
\midrule
Baseline ResNet18    & 300$\times$300 image                                                                     & Rating   distribution                                                                 & EMD loss                                                                     & 0.702 & 0.706 & 80.0\% \\
\cdashline{1-7}
Multi-modal ResNet18 & \begin{tabular}[c]{@{}l@{}}300$\times$300   image\\ Two-hot semantic vector\end{tabular} & Rating   distribution                                                                 & EMD loss                                                                      & 0.709 & 0.712 & 80.2\% \\
\cdashline{1-7}
Multi-task ResNet18  & 300$\times$300 image                                                                     & \begin{tabular}[c]{@{}l@{}}Rating distribution\\ Two-hot semantic vector\end{tabular} & \begin{tabular}[c]{@{}l@{}}Multi-task loss   \end{tabular} & 0.714 & 0.716 & 80.3\% \\
\midrule
KD ResNet18 (proposed)          & 300$\times$300 image                                                                     & Rating   distribution                                                                 & \begin{tabular}[c]{@{}l@{}}KD loss \end{tabular}           & 0.719 & 0.722 & 80.7\% \\
\bottomrule
\end{tabular}
\end{adjustbox}
\label{tbl:discuss_semantics}
\vspace{-6mm}
\end{table*}

\subsection{Performance Variations}
\label{sec:perform}

As mentioned in Sec. \ref{sec:q4}, performance variations can be observed when the same model is evaluated on different train-test splits. As a very relevant topic,  image quality assessment (IQA) methods \cite{zhou2021omnidirectional, zhang2018blind, jiang2017blique, fang2019perceptual, zhang2021fine} typically conduct K-fold cross-validation or take the average or median results evaluated on a number of different random train-test splits (\textit{a.k.a.} sessions) to reduce performance variations as much as possible. However, for the problem of IAA, the adopted datasets by IAA typically have a much greater scale than IQA datasets. For example, the commonly-adopted IQA dataset LIVE-IQA \cite{sheikh2006statistical} contains only 779 images. On the contrary, the commonly-adopted IAA dataset AVA benchmark contains $\sim$255,000 images. This essential difference on dataset scales makes previous works on IAA mostly evaluate the methods only on a fixed train-test split \cite{lu2015deep, mai2016composition, murray2017deep, talebi2018nima, zeng2019unified, zhang2019gated, hosu2019effective, hou2020object} rather than several random splits. However, evaluation on a fixed split potentially causes worries about generalizability (\textit{e.g.}, the method only works on a certain split). Additionally, several versions of train-test splits have been adopted in previous works (as mentioned in Sec. \ref{sec:q4}), which potentially cause problems of unfairness when comparing the results on different train-test splits. Therefore, in this section, we discuss whether the effectiveness of the proposed KD scheme is an outcome of random variations, and whether the effectiveness of the proposed KD scheme only works for a certain train-test split.

\begin{figure}
\centering
\includegraphics[width=0.4\textwidth]{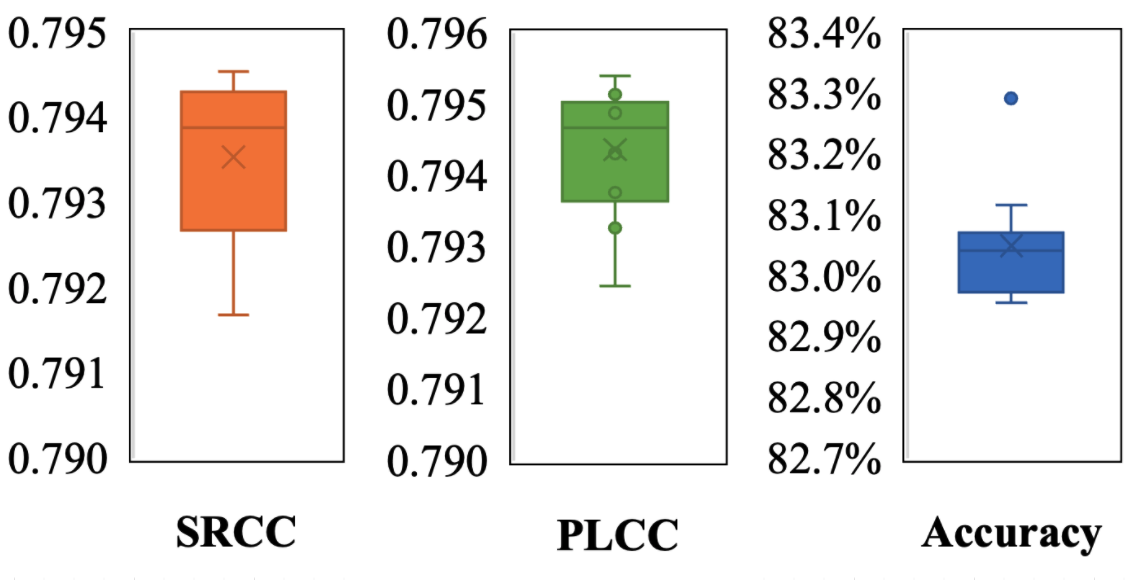}
\caption{Demonstration of performance variations brought by randomness in training. The teacher model has been run for 10 times on the split from \cite{murray2012ava}, and the SRCC, PLCC and accuracy results are plotted in separate box plots. The range for SRCC, PLCC and accuracy are [0.792, 0.795], [0.792, 0.795] and [0.830, 0.833], respectively. Results show the performance variations brought by randomness in training are marginal.}
\label{fig:training_variations}
\vspace{-6mm}
\end{figure}

To begin with, we want to find out the actual range of performance variations brought by different splits. Such variations can be estimated by directly train and test the same model on different splits.
However, due to the randomness in the training of deep learning models, even with the same split, if we separately train a model with the same setting, the results are possibly different. Therefore, we need to separately consider the performance variations brought by train-test split difference alone, and the performance variations brought by randomness in training. That is:
\begin{equation}
    \delta_{exp} = \delta_{split} + \delta_{training},
\end{equation}
where $\delta_{exp}$ is the performance variations observed from experiments that evaluate the same model on different train-test splits, $\delta_{split}$ is the actual performance variation brought by using different train-test split, and $\delta_{training}$ is the performance variation brought by randomness in training. To evaluate the scale of $\delta_{split}$, we first estimate the scale of $\delta_{training}$  by running the proposed teacher model 10 times on the same split. In Fig. \ref{fig:training_variations}, the results show that the variations (difference between the highest and the lowest results) caused by randomness in training are around 0.003 for SRCC, PLCC and accuracy. 

Then we estimate the scale of performance variations brought by using different train-test splits. To this end, we adopt both fixed split evaluation and cross-validation with random splits. For fixed split evaluation, we apply the split from \cite{murray2012ava} and the split from \cite{jin2019ilgnet}, which both left 20,000 images for testing. For cross-validation with random splits, we also remain 20,000 images for testing. Since there are $\sim$255,000 images in total, twelve cross-validation splits with non-overlapping testing sets are prepared. Then we evaluate the results when a single-backbone end-to-end IAA model is cooperated with or without the proposed KD scheme. The results are presented in Fig. \ref{fig:gain_variations}. 
\textcolor{black}{
As the results demonstrated, the proposed KD scheme introduces performance gains for all selected student backbones across all train-test splits. For the ResNet18-based model, the ranges of improvements are [0.011,0.024], [0.010,0.023], [0.5\%,1.7\%] for SRCC, PLCC, and Acc, respectively. For the ResNet50-based model, the ranges of improvements are [0.018,0.026], [0.015,0.041], [0.4\%,1.3\%] for SRCC, PLCC, and Acc, respectively. For the ResNext101-based model, the ranges of improvements are [0.019, 0.038], [0.017,0.036], [0.8\%,1.7\%] for SRCC, PLCC, and Acc, respectively. All improvements are larger than the random variation 0.003, and thus, the improvements on all experimented models on all splits are regarded as significant.}




\begin{figure*}[h]
\centering
\includegraphics[width=0.95\textwidth]{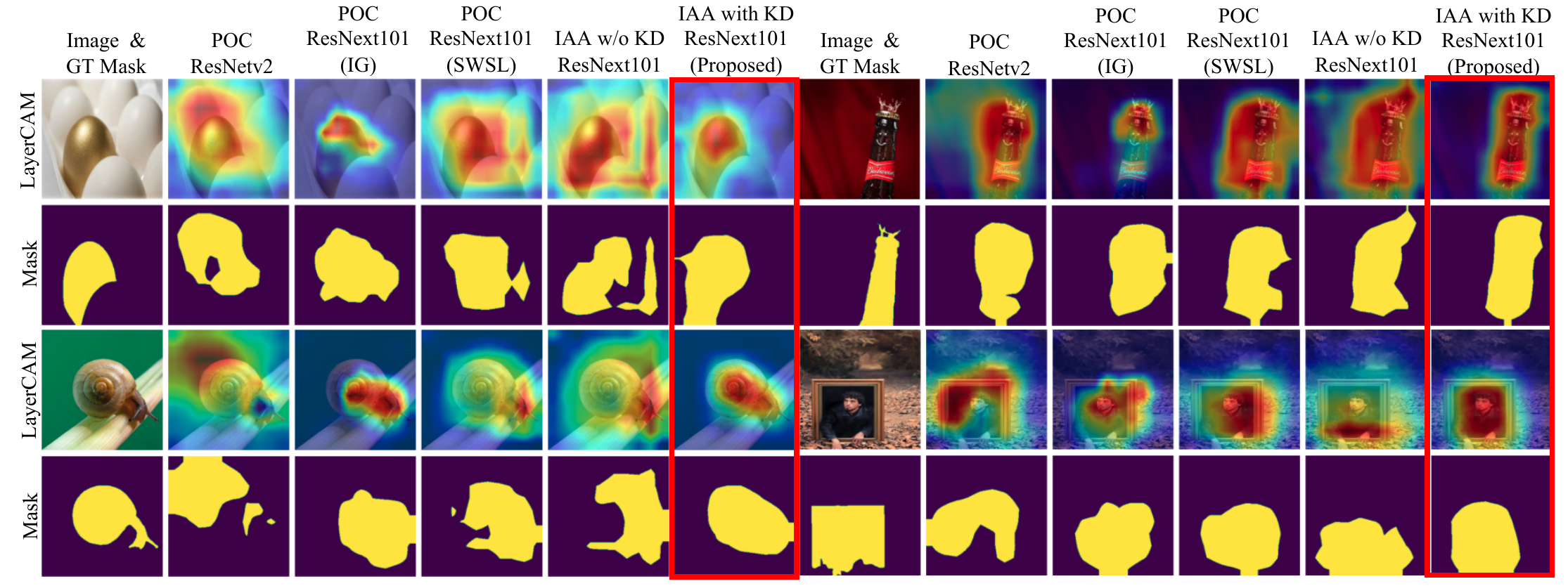}
\caption{Comparison of the capability of different models on extracting the main subjects. We include the selected POC models and IAA models trained with or without (w/o) KD in this comparison. Every two rows form a group of examples, where the top row in each group presents source images and layerCAMs \cite{jiang2021layercam} obtained from the selected models; the bottom row of each group presents GT main subject masks for the source images and masks derived from the corresponding layerCAMs (with 70 percentile of the layerCAMs as the segmentation threshold).  The results show that the IAA model with KD better locates main subjects than POC models and the IAA model without KD, implying teacher knowledge from POC models allows the student model to capture semantic patterns more relevant to subject areas. Further quantitative results are given in Fig. \ref{fig:miou_curve}.} 
\label{fig:layercam}
\vspace{-6mm}
\end{figure*}

\begin{figure}
\includegraphics[width=0.48\textwidth]{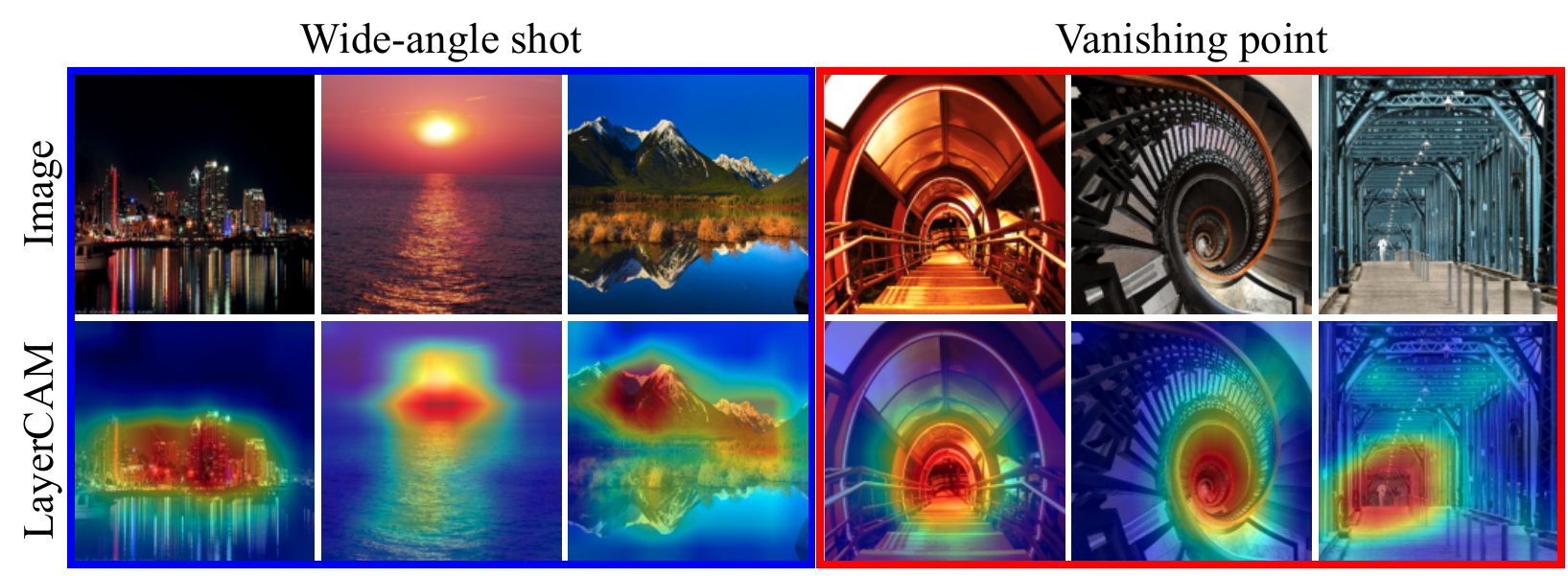}
\caption{LayerCAM examples of cases that do not have a clear main subject. For these cases, the IAA model trained with KD is still able to locate the most appealing parts.} 
\label{fig:sod_exceptions}
\vspace{-6mm}
\end{figure}

\subsection{Comparing Different Strategies for Semantic Guidance}
\label{sec:semantic_guidance}

The proposed KD scheme is designed to provide semantic guidance to the training of an IAA model. Two alternative approaches for introducing semantic guidance are to adopt semantic labels as auxiliary inputs, or train the model to predict semantic labels as an auxiliary task. To compare the effectiveness of the proposed KD scheme with the alternative approaches for introducing semantic guidance, we take the semantic labels provided by the AVA dataset \cite{murray2012ava} (two labels per image and 66 distinct semantic labels in total) for building models that are compared with the model trained with the proposed KD scheme. Therefore, the following four models are built for the comparison:
\subsubsection{Baseline ResNet18} A ResNet18-based end-to-end IAA model is trained with EMD loss subject to aesthetic label prediction supervised by GT aesthetic labels, which follows the same setting as the baselines in our previous experiments in Sec. \ref{sec:q2}. 
\subsubsection{Multi-modal ResNet18} The two semantic labels of each image in the AVA dataset is represented with a two-hot vector. Based on the baseline model, a separate branch constructed with two stacking fully-connected layers is introduced to encode the two-hot semantic vector into semantic features. Then the encoded semantic feature is merged with the visual feature encoded by the ResNet18 for predicting the aesthetic label. Since only aesthetic labels are predicted, the model is trained directly with EMD loss.
\subsubsection{Multi-task ResNet18} Instead of taking the two-hot semantic vector as an auxiliary input, the multi-task ResNet18 setting is designed to predict the two-hot semantic vector. Specifically, based on the baseline model, the multi-task ResNet18 model not only predicts the aesthetic labels according to the visual features from the ResNet18 backbone,  but also predict the two-hot semantic vector with the visual feature. Aesthetic label prediction is learned subject to the EMD loss as previous settings; for the semantic prediction, the model is trained subject to binary cross-entropy (BCE) loss. The multi-task loss is given as:
\begin{equation}
\label{eq:mt_loss}
\mathcal{L_{MT}}({\mathcal{\hat D}_a}, {\mathcal{D}_a}, {\hat v_s}, {v_s}) = EMD({\mathcal{\hat D}_a}, {\mathcal{D}_a}) + BCE({\hat v_s}, {v_s}),
\end{equation}
where ${\mathcal{D}_a}$ denotes a GT aesthetic label (in the form of distribution),  ${v_s}$ represents a GT semantic label (in the form of two-hot vector), ${\mathcal{\hat D}_a}$ denotes a predicted aesthetic label, and 
${\hat v_s}$ denotes a predicted semantic label.
\subsubsection{KD ResNet18} This setting denotes the ResNet18-based IAA model trained with the proposed KD scheme as introduced in Sec. \ref{sec:approach}.

All above-mentioned models are trained with the same setting as Sec. \ref{sec:exp_settings}. The results are presented in Table \ref{tbl:discuss_semantics}. As the table shows, with extra semantic labels, performance gains can be observed from the results of the Multi-modal ResNet18 model or the Multi-task ResNet18. This confirms that semantic guidance can help IAA models to learn how to measure image aesthetics better. However, even with extra semantic labels, the Multi-modal ResNet18 model or the Multi-task ResNet18 are inferior to the setting with KD, this indicates that the proposed KD scheme can more effectively provide semantic guidance than using extra semantic labels, which supports our earlier discussion on weaknesses of using human semantic labels in Sec. \ref{sec:kd_iaa}.

\begin{table*}[]
\caption{Comparison between extra computational costs brought by the proposed KD scheme and the computational costs for training or testing an end-to-end IAA model. }
\renewcommand\arraystretch{1.1} 
\begin{adjustbox}{width=2\columnwidth,center}
\begin{tabular}{l|l|l|l|ll|l}
\toprule
Setting & Remark                                  & Model                                               & Input size    & \begin{tabular}[c]{@{}l@{}}Cost per input  \\ (in FLOPs(G))\end{tabular} & \# Epochs & \begin{tabular}[c]{@{}l@{}}Cost per input $\times$ \# Epochs\\ (in FLOPs(G))\end{tabular} \\
\midrule
1       & \multirow{3}{*}{Feature extraction with POC models} & ResNeXt101 (SWSL)                                   & (3, 640, 640) & 134.7                                                                 & 1         & 134.7                                                                          \\
2       &                                         & ResNeXt101 (IG)                                     & (3, 640, 640) & 1280                                                                   & 1         & 1280.0                                                                            \\
3       &                                         & ResNet-v2                                            & (3, 640, 640) & 1525.8                                                                & 1         & 1525.8                                                                         \\
\cdashline{2-7}
4       & Train the knowledge distiller           & in Fig.\ref{fig:distiller} & (1, 23424)    & 0.3                                                                    & 12        & 3.6                                                                            \\
5       & Distill teacher knowledge               & in Fig.\ref{fig:distiller} & (1, 23424)    & 0.1                                                                    & 1         & 0.1                                                                             \\
\midrule
\multicolumn{6}{r|}{Total extra cost per input}                                                                                                                                                               & 2944.2                                                                         \\
\midrule
6       & Train the end-to-end IAA model with KD                 & ResNeXt101 (SWSL)                                   & (3, 300, 300) & 91.8                                                                  & 12        & 1101.2                                                                         \\
7       & Test the end-to-end IAA model with KD                  & ResNeXt101 (SWSL)                                   & (3, 300, 300) & 30.6                                                                  & 1         & 30.6                                                                           \\
\cdashline{2-7}
8       & Train the end-to-end IAA model w/o KD                 & ResNeXt101 (SWSL)                                   & (3, 300, 300) & 91.8                                                                  & 12        & 1101.2                                                                         \\
9      & Test the end-to-end IAA model w/o KD                 & ResNeXt101 (SWSL)                                   & (3, 300, 300) & 30.6                                                                  & 1         & 30.6                                                                          \\
\bottomrule
\end{tabular}
\end{adjustbox}
\label{tab:cost}
\vspace{-6mm}
\end{table*}

\subsection{Capability of Extracting the Main Subjects}
\label{sec:aesthetic_sod}

Since it is a widely accepted opinion that an aesthetically-pleasant image should be able to lead the attention of the observer to the main subject \cite{luo2008photo}, semantic patterns for aesthetic features should be relevant to the subject area.
Therefore, a more explainable way for measuring the performance of a deep IAA model is to evaluate the capability of a deep IAA model on extracting main subjects from aesthetically-pleasant images. To this end, we collect a number of aesthetically-pleasant images and manually label their main subjects. Then for different IAA models, we can derive the predicted main subjects from their class activation maps and compare the accuracy of extracting main subjects across different models.

Specifically, we collect those aesthetically-pleasant images from the AVA testing set under the split from \cite{murray2012ava}. Images with average scores greater than 6.5 are regarded as aesthetically-pleasant images. Then main subjects are firstly detected with a salient object detection model \cite{qin2019basnet}. We find that some types of images, such as landscapes, may not have a clear main subject. For these cases, it may be hard to conclude a definite salient object detection result. Therefore, we firstly select images whose detected main subjects occupy 10\%$\sim$30\% of its area. Then we manually remove images with poor salient object detection results. Finally, 504 images with salient object masks are prepared, where the salient object masks are regarded as GT labels for the main subjects. Note that for those removed cases, the IAA model trained with the proposed KD scheme is still able to locate their most appealing part as demonstrated in Fig. \ref{fig:sod_exceptions}. While these cases do not have a commonly-accepted guideline to outline their main subjects, and therefore, these cases are not included in the evaluation of this section. 

The model trained with the proposed KD scheme is then compared to the version trained without KD and the selected POC models in terms of locating the main subjects. Some selected cases are given in Fig. \ref{fig:layercam}. We firstly extract layerCAMs \cite{jiang2021layercam} from the selected models. Then masks for the main subjects are derived from the layerCAMs by setting the 70 percentile of the layerCAM as the threshold for segmenting the regions of the main subjects. We can observe from  \figurename~\ref{fig:layercam} that: 1) different POC models tend to have different focuses, implying different POC models have sensitivities to different semantic patterns; 2) the IAA model without KD captures semantic patterns less relevant to the subject area, which supports our claims that aesthetic labels are too abstract to guide the deep model to learn about semantic patterns for constructing aesthetic features; 3) the IAA model with KD supervised by knowledge on semantic patterns tend to have a better focus on subject areas than the version without KD, implying that KD allows the student model to capture semantic patterns more relevant to the subject area; 4) the focused subject areas of the IAA model with KD are further improved from those of the POC models, which implies that the knowledge distiller can build aesthetic features upon GSFs.

\begin{figure}
\includegraphics[width=0.48\textwidth]{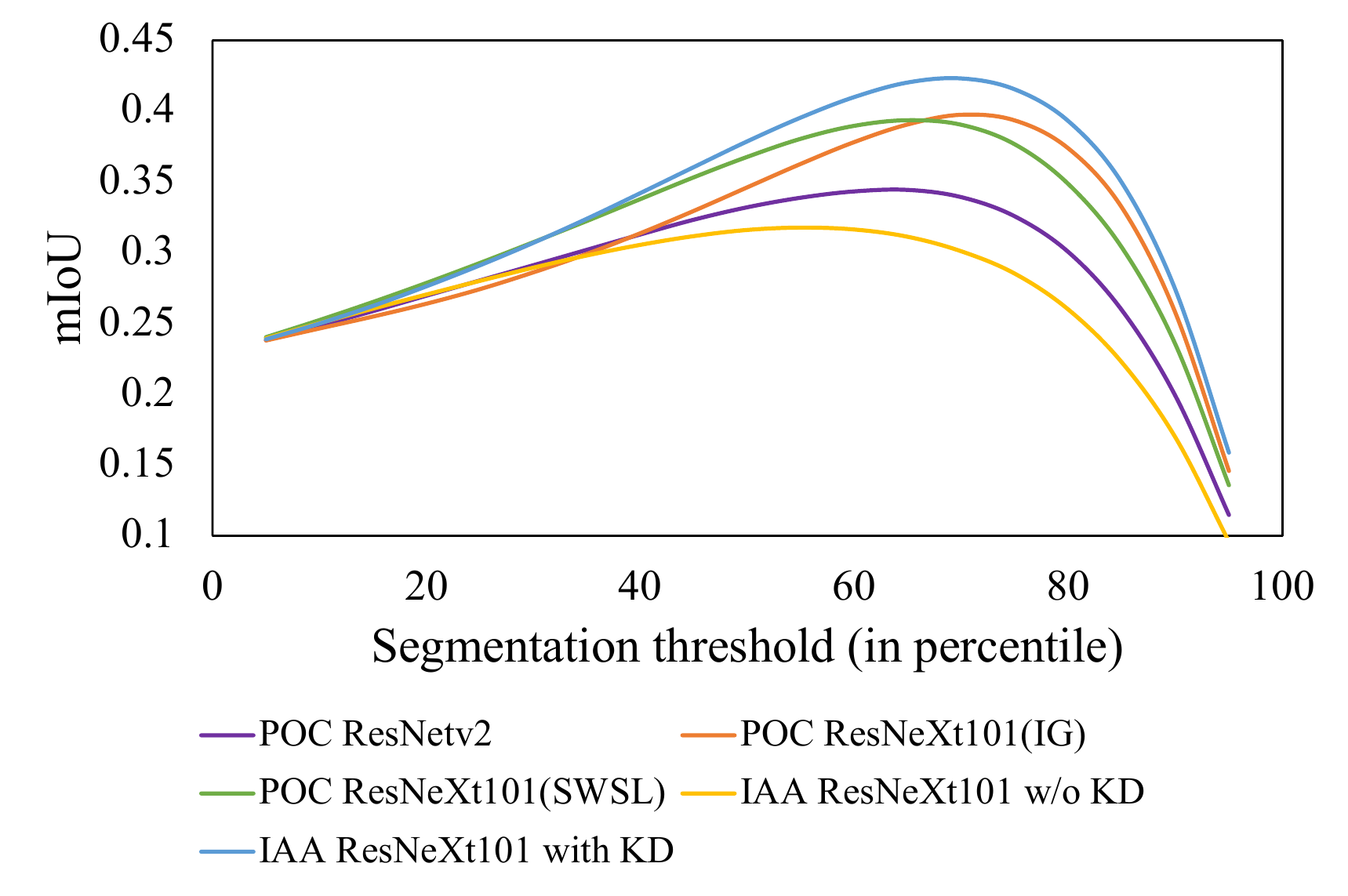}
\caption{Curves for depicting mIoU varying by segmentation thresholds from 5 to 95 percentile. Different curves present the performance of different models on extracting the main subjects on aesthetically-pleasant images in terms of mean intersection-over-union (mIoU). We include the selected POC models and IAA models trained with or without (w/o) KD in this comparison. Higher mIoU means better performances on extracting main subjects on our collected dataset. }
\label{fig:miou_curve}
\vspace{-6mm}
\end{figure}

Besides qualitatively analyzing the impact of teacher knowledge on the student model, to confirm the judgments in qualitative analysis, we further quantitatively evaluate the performances of the selected models on extracting main subjects with the collected dataset. To this end, we measure mean intersection-over-union (mIoU) between the masks derived from layerCAMs and the GT masks. 
Since the mIoU results are varied subject to different segmentation thresholds, we plot the change of mIoU subject to segmentation thresholds from 5 to 95 percentile of the layerCAMs. 
According to Fig. \ref{fig:miou_curve}, from best to worst, the performance on main subject detection can be ranked as: IAA ResNeXt101 with KD $>$ selected POC models $>$ IAA ResNeXt101 without KD. This further confirms our judgments concluded in the qualitative analysis. 


\subsection{Cost and Benefit Analysis} 

The extra cost of the proposed KD scheme is brought by: 1) extracting GSFs from multiple POC models; 2) training the knowledge distiller; 3) distilling teacher knowledge from GSFs; 4) computing losses between teacher aesthetic features and student aesthetic features when training the student model. To compare the scale of these extra computational costs to the computational cost for training a baseline IAA model, the costs are estimated and listed in Table \ref{tab:cost}. Specifically, the scale of the training cost is measured as three times of the testing costs for the same model with the same input \cite{flopsestimation}. Since all the above-mentioned four parts that bring extra computational costs can be conducted separately, the overall extra computational costs are computed by linear combination. In Table \ref{tab:cost}, to show the total extra cost brought by the proposed KD scheme, we present the costs for feature extraction with POC models (Setting 1, 2, 3), training the knowledge distiller (Setting 4), and distill teacher knowledge with the knowledge distiller (Setting 5). Compared to the forward-propagation or backward-propagation via the models, the computational costs for loss computations are negligible, and therefore, costs for loss computations are not listed in Table \ref{tab:cost}. Therefore, the total extra cost per input is the sum of the costs brought by feature extraction with POC models, training the knowledge distiller, and distilling teacher knowledge with the knowledge distiller. As to the cost for training the end-to-end IAA model with or without KD (Setting 6, 8), their computational costs are the same since the negligible cost for computing the KD loss is the only difference that brings the extra cost for the setting with KD. For the cost for testing the end-to-end IAA models with or without KD (Setting 7, 9), the costs are the same since their settings are the same in the testing phase. 

As it can be seen from Table \ref{tab:cost}, the total extra cost is 2944.2 FLOPs(G) while the training cost for an end-to-end IAA model is 1101.2 FLOPs(G) (Setting 6, 8). This means that the total extra cost is $\sim$1.7 times higher than training an end-to-end IAA model. Do note that the total training cost (Setting 6, 8) is a multiple of the number of epochs, while our setting adopts a rather fewer number of epochs in training (\textit{i.e.,} 12 epochs), which makes the extra cost seemingly higher. While it is common to use a few tens of epochs in training an end-to-end IAA model. For example, Li \textit{et al.}'s method \cite{li2020personality} adopts 50 epochs in training. If our model is also trained for a few tens of epochs, the extra cost will be less than the training costs for an end-to-end model. On the contrary, among all three parts within the extra cost (Setting 1$\sim$5), only the cost brought by training the knowledge distiller is a multiple of the number of epochs (Setting 4), and other parts of the extra cost will stay constant. Since the cost brought by training the knowledge distiller is on a scale of 1, increasing the number of epochs for training the knowledge distiller does not significantly increase the overall extra cost. With the extra cost, the model performance can be effectively improved by 4.8\% (in Table \ref{tab:expq3}), and such an improvement is generalizable (in Fig. \ref{fig:gain_variations}). Therefore, we believe that the proposed KD scheme is a rather cost-effective way for improving the performance of an end-to-end IAA model. 
\begin{table}[]
\caption{Comparison of training with GT (with mixed loss or mixed label) or without GT (with teacher prediction only) along with teacher predictions for supervision (evaluated on train-test split from  \cite{murray2012ava}).}
\renewcommand\arraystretch{1.1} 
\begin{adjustbox}{width=0.8\columnwidth,center}
\begin{tabular}{l|lll}
\toprule
\multicolumn{1}{c|}{Setting} & \multicolumn{1}{c}{SRCC} & \multicolumn{1}{c}{PLCC} & \multicolumn{1}{c}{Acc} \\
\midrule
with teacher prediction       & 0.770                    & 0.770                     & 82.1\%                   \\
\cdashline{1-4}
with mixed loss               & 0.773                    & 0.773                    & 82.2\%                   \\
with mixed label              & 0.769                    & 0.769                    & 82.0\%                    \\
\bottomrule
\end{tabular}
\end{adjustbox}
\label{tbl:gtwithteacher}
\vspace{-6mm}
\end{table}

\begin{table}[]
\caption{Results for ablation studies on tasks besides distribution prediction (evaluated on train-test split from  \cite{murray2012ava}).}
\renewcommand\arraystretch{1.1} 
\begin{adjustbox}{width=0.95\columnwidth,center}
\begin{tabular}{c|c|c|ccc}
\toprule
Dataset & Task & Setting & SRCC & PLCC & Acc \\
\toprule
\multirow{3}{*}{CUHK-PQ} & \multirow{3}{*}{Classification} & Teacher  & -      & -     & 96.9\% \\
                         \cdashline{3-6}
                         &                                 & w/o KD & -      & -     & 96.2\% \\
                         &                                 & with KD  & -      & -     & 96.9\% \\
\midrule
\multirow{3}{*}{AVA}     & \multirow{3}{*}{Classification} & Teacher  & -      & -     & 82.8\% \\
                         \cdashline{3-6}
                         &                                 & w/o KD & -      & -     & 80.0\% \\
                         &                                 & with KD  & -      & -     & 81.7\% \\
\midrule
\multirow{3}{*}{AVA}     & \multirow{3}{*}{Regression}     & Teacher  & 0.784  & 0.785 & 82.3\% \\
                         \cdashline{3-6}
                         &                                 & w/o KD & 0.687  & 0.689 & 78.0\% \\
                         &                                 & with KD  & 0.707  & 0.693 & 78.9\% \\
\bottomrule
\end{tabular}
\end{adjustbox}
\label{tbl:abl_cls_rgrs}
\vspace{-6mm}
\end{table}

\vspace{-4mm}

\subsection{Discussion on Cause of Improvements brought by KD}

\textcolor{black}{
A better IAA model requires more discriminative aesthetic features, and more discriminative aesthetic features require more diverse semantic patterns. 
}

\textcolor{black}{
For the teacher model, using POC model with larger sizes, using POC model with more training data, or combining GSFs from different POC models can significantly improve the teacher models' performance (Table \ref{tbl:GSF_resnet} and Table \ref{tbl:GSF_resnext}). The main reason is that all above-mentioned approaches can provide more diverse semantic patterns that potentially provides more relevant patterns for constructing aesthetic features covering more contents. As a result, more discriminative aesthetic features further lead to performance gains.
}

\textcolor{black}{
For the student model, the performance improvement brought by KD can be explained by the semantic patterns learned from the teacher. 
As we experimentally show in Sec. \ref{sec:exp1} and Sec. \ref{sec:q2}, both the teacher model and the student model trained merely with aesthetic labels construct aesthetic features from semantic patterns captured by their own pretrained backbones. 
Since the selected POC models for the teacher model can capture more semantic patterns than the student's backbone, the teacher aesthetic features are expected to cover more contents than the student aesthetic features.
Thus, teacher aesthetic features can be used for extra supervision that guides the student model to capture more relevant semantic patterns for constructing aesthetic features covering more contents, which leads to higher IAA performance. Besides ablation studies on IAA in Table \ref{tab:expq2}, results on detecting the subject areas (Fig. \ref{fig:layercam} and Fig. \ref{fig:miou_curve}) also imply that the IAA model learns to capture more IAA-relevant semantic patterns (related to the subject areas). 
}

\subsection{Variants of KD Loss}

\textcolor{black}{
In this section, we further investigate some variants of the proposed KD loss along with the proposed method.
\subsubsection{Combine GT with teacher predictions} We first investigate whether combining GT with teacher predictions will further improve the student model's performance. To this end, we have set up two settings to supervise the student model with the mixture of GT and teacher predictions:
\begin{itemize}
    \item  {\textbf{Mixed loss}}: GT and teacher predictions are separately used for loss computation as follows:
    \begin{equation}
        \begin{aligned}
        \mathcal{L}_1({\mathcal{\hat D}_t}, {\mathcal{\hat D}_s}, {\mathcal{D}_{GT}}, {f_t}, {f_s}) = \frac{1}{2} EMD({\mathcal{\hat D}_t}, {\mathcal{\hat D}_s})  \\
        + \frac{1}{2} EMD({\mathcal{D}_{GT}}, {\mathcal{\hat D}_s}) + MSE({f_t}, {f_s}),
        \end{aligned}
    \end{equation}
    where ${\mathcal{\hat D}_t}, {f_t}$ denote the teacher prediction and the teacher aesthetic feature, ${\mathcal{\hat D}_s}, {f_s}$ denote the student prediction and the student aesthetic feature, ${\mathcal{D}_{GT}}$ denotes GT, $EMD(\cdot)$ and $MSE(\cdot)$ refer to EMD loss and mean squared error (MSE) loss respectively. 
    \item  {\textbf{Mixed loss}}: GT and teacher predictions are linearly combined before loss computation:
    \begin{equation}
        \begin{aligned}
        \mathcal{L}_2({\mathcal{\hat D}_t}, {\mathcal{\hat D}_s}, {\mathcal{D}_{GT}}, {f_t}, {f_s}) =  EMD({\frac{1}{2}\mathcal{D}_{GT}} +  \frac{1}{2}{\mathcal{\hat D}_t}, {\mathcal{\hat D}_s}) \\
        + MSE({f_t}, {f_s}).
        \end{aligned}
        \end{equation}
\end{itemize}
The above-mentioned two settings are compared with the results of supervision with only teacher predictions. All models are based on ResNeXt101 (SWSL). The experimental results are given in Table \ref{tbl:gtwithteacher}. As the results show, both settings with a mixture of GT and teacher predictions as supervision do not significantly outperform the setting merely with teacher predictions as supervision, considering a 0.003 performance variations as discussed in Sec. \ref{sec:perform}. We interpret the results as follows. Compared to directly supervising the student model with GT, supervision with teacher predictions provide easier solutions to the student to map from aesthetic features to aesthetic predictions. Even trained with easier supervisions, there is still a performance gap between the teacher model and the student model. This means even we mix the GT with the teacher predictions for supervision, it is less likely that the student model will able to learn more information from GT.
}

\textcolor{black}{
\subsubsection{Generalized to other IAA tasks} Since the distribution labels are not available on some IAA datasets, it would be better to verify the effectiveness of the proposed method when binary labels or MOSs are used as the GT. To this end, we adjust the KD loss to the cases when only binary labels or MOSs are available. 
\begin{itemize}
    \item For the case of binary classification, the first term can be replaced with a binary cross-entropy (BCE) loss:
\begin{equation}
    \begin{aligned}
    \mathcal{L_{KD}}({\hat y_t}, {\hat y}_s, {f_t}, {f_s}) =  BCE({\hat y}_t, {\hat y}_s) + MSE({f_t}, {f_s}),
    \end{aligned}
\end{equation}
where ${\hat y_t}\in[0,1], {f_t}$ denote the teacher prediction and the teacher aesthetic feature, ${\hat y_s}\in[0,1], {f_s}$ denote the student prediction and the student aesthetic feature,  $BCE(\cdot)$ and $MSE(\cdot)$ refer to BCE loss and mean squared error (MSE) loss respectively. 
    \item For the case of score regression, the first term can be replaced with a MSE loss:
\begin{equation}
    \begin{aligned}
    \mathcal{L_{KD}}({\hat y_t}, {\hat y}_s, {f_t}, {f_s}) =  MSE({\hat y}_t, {\hat y}_s) + MSE({f_t}, {f_s}),
    \end{aligned}
\end{equation}
where ${\hat y_t}, {f_t}$ denote the teacher prediction and the teacher aesthetic feature, ${\hat y_s}, {f_s}$ denote the student prediction and the student aesthetic feature,  $MSE(\cdot)$ and $MSE(\cdot)$ refer to MSE loss and mean squared error (MSE) loss respectively. 
\end{itemize}
We have conducted experiments on both CUHK-PQ dataset \cite{luo2011content} (binary classification) and AVA dataset (binary classification and score regression) with above-mentioned losses. The results are given in Table \ref{tbl:abl_cls_rgrs}. The results show that the proposed KD scheme is still effective in the two other cases of IAA.
}


\section{Conclusion}



\textcolor{black}{
In this paper, we have focused on the problem of abstractness of aesthetic labels. 
On the one hand, during inference, the IAA model is required to relate various distinct semantic patterns to the same aesthetic label. On the other hand, when training, it would be hard for the IAA model to learn to distinguish different semantic patterns merely with the supervision from aesthetic labels.
When the supervision was merely provided by aesthetic labels, experimental results (Table \ref{tab:expq2}) have implied that an IAA model mostly learned to construct aesthetic features from semantic patterns already-known to its pre-trained backbone, instead of learning new semantic patterns for more discriminative aesthetic features.
Therefore, an IAA model can be improved by providing semantic guidance in training, so that the IAA model can learn extra semantic patterns for more discriminative aesthetic features. 
}

\textcolor{black}{
The proposed method is inspired by the observation that different POC models tend to capture different sets of semantic patterns (Fig. \ref{fig:layercam}).
And therefore, an IAA model that captures less diverse semantic patterns can learn from these POC models to capture more diverse semantic patterns for more discriminative aesthetic features covering more contents. However, since knowledge from POC models may not directly applicable to IAA (Fig. \ref{fig:GSF_vs_AF}, Fig. \ref{fig:redundant}, and Table \ref{tbl:GSF_vs_AF}), we have proposed to train a separate knowledge distiller to take relevant parts from the knowledge of POC models. Thus, a single-backbone end-to-end model can be trained for IAA with semantic guidance from multiple POC models (Fig. \ref{fig:overview}). Since the selected POC models are deeper or trained with more data, extra semantic patterns are learned by the student for more discriminative aesthetic features (Table \ref{tbl:GSF_resnet}, Table \ref{tbl:GSF_resnext}, Table \ref{tab:stacked_features}).
}

\textcolor{black}{
Extensive experiments showed that: 1) the proposed KD scheme enables the student IAA model to capture more IAA-relevant semantic patterns for building more discriminative aesthetic features (Table \ref{tab:expq2}, \ref{tab:expq22}, \ref{tab:subsets}, Fig. \ref{fig:layercam}, \ref{fig:miou_curve}); 2) the proposed KD scheme also allows the student IAA model to adapt to smaller-sized inputs (Table \ref{tab:expq3}, \ref{tab:expq23}); 3) the proposed KD scheme is generalizable across different train-test splits (Fig. \ref{fig:training_variations}, \ref{fig:gain_variations}); 4) the proposed KD scheme is a rather cost-effective way for improving the performance of an IAA model (Table \ref{tab:cost}); 5) the proposed IAA method significantly outperforms 10 previous relevant methods (Table \ref{tab:expq4_official}, \ref{tab:expq4_ilgnet}); 6) the proposed approach can be adapted to binary classification and regression scenarios of IAA while keeping  its effectiveness (Table \ref{tbl:abl_cls_rgrs}).
}


%





\ifCLASSOPTIONcaptionsoff
  \newpage
\fi

\vspace{-3mm}
\bibliographystyle{IEEEtran}
\bibliography{IEEEexample}







\end{document}